\documentclass[showpacs,preprintnumbers,amsmath,amssymb]{revtex4}

\usepackage{graphicx}
\usepackage{dcolumn}
\usepackage{bm}

\newcommand{\bq}{\begin{equation}}
\newcommand{\eq}{\end{equation}}
\newcommand{\bqa}{\begin{eqnarray}}
\newcommand{\eqa}{\end{eqnarray}}
\newcommand{\nn}{\nonumber \\}

\def\be     {\begin{equation}}
\def\ee     {\end{equation}}
\def\bea        {\begin{eqnarray}}
\def\eea        {\end{eqnarray}}
\def\bnn    {\begin{eqnarray*}}
\def\enn    {\end{eqnarray*}}

\begin{document}

\title{Emergent geometry in recursive renormalization group transformations}
\author{Ki-Seok Kim}
\affiliation{Department of Physics, POSTECH, Pohang, Gyeongbuk 790-784, Korea}
\date{\today}

\begin{abstract}
Holographic duality conjecture has been proposed to be a novel non-perturbative theoretical framework for the description of strongly correlated electrons. However, the duality transformation is not specified to cause ambiguity in the application of this theoretical machinery to condensed matter physics. In this study, we propose a prescription for the holographic duality transformation. Based on recursive renormalization group (RG) transformations, we obtain an effective field theory, which manifests the RG flow of an effective action through the introduction of an extra dimension. Resorting to this prescription, we show that RG equations of all coupling constants are reformulated as emergent geometry with an extra dimension. We claim that the present prescription serves as microscopic foundation for the application of the holographic duality conjecture to condensed matter physics.
\end{abstract}


\maketitle

\section{Introduction}

Holographic duality conjecture \cite{Holographic_Duality_I,Holographic_Duality_II,Holographic_Duality_III,Holographic_Duality_IV,Holographic_Duality_V, Holographic_Duality_VI,Holographic_Duality_VII} states that a $(D+1)-$dimensional quantum gravity theory is a dual hologram of a $D-$dimensional quantum field theory, where $D$ is a spacetime dimension. The emergent extra dimension of the quantum gravity theory is claimed to play the role of an energy scale for the renormalization group (RG) transformation of the quantum field theory. In other words, the bulk quantum gravity theory manifests an RG flow of the boundary quantum field theory through the evolution of geometry along the extra-dimensional space. This emergent quantum gravity theory describes the dynamics of conserved currents of the quantum field theory in terms of dual field variables such as metric tensor and U(1) gauge fields. It turns out that quantum fluctuations of such dual field variables are suppressed when the number of ``local degrees of freedom" is ``infinite". Taking the so called large $N$ limit, the quantum gravity theory becomes classical. This classical gravity theory is regarded as a novel mean-field theory which takes into account renormalization effects in a non-perturbative way beyond the conventional mean-field theory.

So far, there is no consensus for the holographic duality transformation that maps the boundary quantum field theory into the bulk quantum gravity theory \cite{RG_Holography_I,RG_Holography_II,RG_Holography_III,RG_Holography_IV,RG_Holography_V,RG_Holography_VI,RG_Holography_VII,RG_Holography_VIII,
RG_Holography_IX,RG_Holography_X,RG_Holography_XI,RG_Holography_XII,RG_Holography_XIII,RG_Holography_XIV,RG_Holography_XV,RG_Holography_XVI,
RG_Holography_XVII,RG_Holography_XVIII,RG_Holography_XIX,RG_Holography_XX,RG_Holography_XXI,SungSik_Holography_I,SungSik_Holography_II,
SungSik_Holography_III,Holographic_Description_Entanglement_Entropy,Holographic_Description_Kondo_Effect,Holographic_Description_Einstein_Klein_Gordon,
Holographic_Description_Einstein_Maxwell,Holographic_Description_RG_GR}. In other words, it is not clear how to relate the RG equations of the coupling constants in the quantum field theory with the evolution of the geometry along the extra-dimensional space in the quantum gravity theory. We believe that this ambiguity gives rise to an obstruction in the application of this non-perturbative theoretical framework to condensed matter physics.

Recently, we proposed a prescription for the holographic duality transformation based on the Wilsonian scheme of RG transformations \cite{Holographic_Description_Entanglement_Entropy,Holographic_Description_Kondo_Effect,Holographic_Description_Einstein_Klein_Gordon,
Holographic_Description_Einstein_Maxwell,Holographic_Description_RG_GR}. Here, we put a $D-$dimensional quantum field theory on a $D-$dimensional curved spacetime just formally although the quantum field theory is originally written in a $D-$dimensional Euclidean spacetime. In addition, we perform the Hubbard-Stratonovich transformation for effective interactions that introduces a collective dual field variable into the quantum field theory in a standard fashion. Then, we apply the Wilsonian RG transformation to this effective field theory. Separating all dynamical field variables into slow and fast degrees of freedom, we perform the path integral with respect to the fast modes of both original matter fields and collective dual field variables. As a result, we obtain an effective field theory in terms of the slow modes, where their effective interactions are renormalized by quantum fluctuations of both heavy degrees of freedom. An essential point is that this Wilsonian RG transformation generates novel effective interactions for low-energy quantum fluctuations, which cannot be fitted into the original form of the resulting effective action by renormalizations of interaction parameters. Here, renormalization effects of effective interactions are given by quantum fluctuations of high-energy original matter fields and described by updating the metric tensor that returns the effective action into its original form. Since these newly generated effective interactions originate from quantum fluctuations of high-energy collective dual field variables, an idea to complete this Wilsonian RG transformation is to perform the Hubbard-Stratonovich transformation once again, which introduces the second collective dual field variable into the resulting renormalized effective action. Now, the effective field theory of the original form is described by the renormalized metric tensor and the second order parameter field.

To compare this RG procedure with the conventional mean-field theory, one may take the large $N$ limit, where $N$ is the flavor number of original quantum fields. Then, quantum fluctuations of such two collective order parameter fields are suppressed and their dynamics are described by two coupled mean-field equations. These coupled mean-field equations turn out to describe that the first order parameter evolves into the second one under the renormalization, where interaction parameters are renormalized by the RG transformation. Since these mean-field equations describe the RG flow of the order parameter field, it is natural to continue the RG transformation in a recursive way. Then, renormalization effects are accumulated and described by the RG flow of the metric tensor. One can reformulate the iteration steps of the RG transformations as an extra-dimensional space. As a result, the RG flow of the metric tensor manifests in the level of the effective action. Furthermore, the RG flow of the collective dual field also manifests through the emergent extra-dimensional space, which may be identified with the Callan-Symanzik equation \cite{RG_Textbook} of a one-particle Green's function. This recursive RG framework takes into account renormalization effects of both interaction vertices and order parameter fields in a self-consistent way, regarded to be a non-perturbative theoretical framework beyond the conventional mean-field theory.

In this study we show equivalence between the RG flow of the metric tensor and the RG $\beta-$functions of interaction vertices based on the above prescription for the holographic duality conjecture. More precisely, we prove the equivalence between the partition function of Eq. (\ref{Partition_Function_Metric_RG_Flow}) and that of Eq. (\ref{Partition_Function_Coupling_RG_Flow}) up to a normalization constant. Here, we focus on the case of one space dimension for an explicit demonstration. Based on this correspondence, we confirm the emergence of AdS$_{3}$ metric at the quantum critical point of interacting scalar bosons.

\section{Overview: Dual holographic effective field theories based on recursive renormalization group transformations} \label{Overview}

%
%

We define our problem precisely. We consider an effective field theory in $D-$spacetime dimensions \cite{Coupling_Scalarfields_Riccicurvature}
\bqa && Z = \int D \phi_{\alpha}(x) \exp\Big[ - \int d^{D} x \sqrt{g_{B}(x)} \Big\{ g_{B}^{\mu\nu}(x) [\partial_{\mu} \phi_{\alpha}(x)] [\partial_{\nu} \phi_{\alpha}(x)] + m^{2} \phi_{\alpha}^{2}(x) + \frac{u}{2 N} \phi_{\alpha}^{2}(x) \phi_{\beta}^{2}(x) \Big\} \Big] . \label{Partition_Function_Metric} \eqa
Here, $\phi_{\alpha}(x)$ is a real scalar field with a flavor index $\alpha$ that runs from $1$ to $N$. $g_{B}^{\mu\nu}(x)$ is a background metric tensor, given by $g_{B}^{\tau\tau}(x) = 1$ and $g_{B}^{ij}(x) = v_{\phi}^{2} \delta_{ij}$ with $g_{B}^{\tau j}(x) = g_{B}^{i \tau}(x) = 0$, where $i$ and $j$ run from $1$ to $D-1$. $m$ is the mass of this scalar field and $u$ is the strength of self-interactions between these scalar bosons. We emphasize that the metric tensor has been introduced into this effective field theory just formally. In this study we focus on $D = 2$ and $N = 1$, which corresponds to an effective field theory for the transverse-field Ising model \cite{Kadanoff_RG}.

Recently, we applied recursive Wilsonian RG transformations to this interacting boson model \cite{Holographic_Description_Einstein_Klein_Gordon} as discussed in the introduction. First, we perform the Hubbard-Stratonivich transformation to decompose the self-interaction term in terms of a collective dual scalar field $\varphi(x)$. Second, we separate both scalar fields of $\phi_{\alpha}(x)$ and $\varphi(x)$ into their slow and fast degrees of freedom. Third, we integrate over all the fast modes and obtain an effective field theory in terms of the slow modes with renormalized interactions. To return this expression into the original form of the effective field theory, we update the metric tensor with rescaling. However, it turns out that quantum fluctuations of the heavy dual scalar bosons result in effective self-interactions. Performing the Hubbard-Stratonovich transformation for this newly generated self-interaction term as the fourth step, we obtain an effective field theory with the second dual scalar field. Fifth, reshuffling several terms and updating the metric tensor with rescaling, we complete the first Wilsonian RG transformation and find a recursive expression for the effective field theory. Sixth, we repeat this Wilsonian RG transformation and complete a recursive expression of the effective field theory. Finally, we take a continuum approximation for the number of recursive Wilsonian RG transformations and reformulate the recursive expression of the effective field theory with the introduction of an extra-dimensional space. As a result, the RG flow of the effective field theory manifests in the resulting partition function through the extra-dimensional space. It turns out that this recursive Wilsonian RG procedure is completely straightforward and easily implemented \cite{Holographic_Description_Entanglement_Entropy,Holographic_Description_Kondo_Effect,Holographic_Description_Einstein_Klein_Gordon,
Holographic_Description_Einstein_Maxwell,Holographic_Description_RG_GR}.

The effective partition function is given by \cite{Holographic_Description_Einstein_Klein_Gordon}
\bqa && Z = \int D \phi_{\alpha}(x) D \varphi(x,z) D g_{\mu\nu}(x,z) \delta\Big(g^{\mu\nu}(x,0) - g_{B}^{\mu\nu}(x)\Big) \nn && \delta\Big\{\partial_{z} g^{\mu\nu}(x,z) - g^{\mu\nu'}(x,z) \Big(\partial_{\nu'} \partial_{\mu'} G_{xx'}[g_{\mu\nu}(x,z),\varphi(x,z)]\Big)_{x' \rightarrow x} g^{\mu'\nu}(x,z) \Big\} \nn && \exp\Big[ - \int d^{D} x \sqrt{g(x,z_{f})} \Big\{ g^{\mu\nu}(x,z_{f}) [\partial_{\mu} \phi_{\alpha}(x)] [\partial_{\nu} \phi_{\alpha}(x)] + [m^{2} - i \varphi(x,z_{f})] \phi_{\alpha}^{2}(x) \Big\} \nn && - N \int d^{D} x \sqrt{g(x,0)} \Big\{ \frac{1}{2u} [\varphi(x,0)]^{2} \Big\} \nn && - N \int_{0}^{z_{f}} d z \int d^{D} x \sqrt{g(x,z)} \Big\{ \frac{1}{2u} [\partial_{z} \varphi(x,z)]^{2} + \frac{\mathcal{C}_{\varphi}}{2} g^{\mu\nu}(x,z) [\partial_{\mu} \varphi(x,z)] [\partial_{\nu} \varphi(x,z)] + \frac{1}{2 \kappa} \Big( R(x,z) - 2 \Lambda \Big) \Big\} \Big] . \label{Partition_Function_Metric_RG_Flow} \eqa
We emphasize that the RG flow of the effective action manifests through the emergence of an extra-dimensional space although this partition function is exactly identical to the original partition function of Eq. (\ref{Partition_Function_Metric}) up to a normalization constant. As discussed in the introduction and above, the metric tensor evolves by this RG transformation, given by $\partial_{z} g^{\mu\nu}(x,z) = g^{\mu\nu'}(x,z) \Big(\partial_{\nu'} \partial_{\mu'} G_{xx'}[g_{\mu\nu}(x,z),\varphi(x,z)]\Big)_{x' \rightarrow x} g^{\mu'\nu}(x,z)$ inside the $\delta-$function constraint. Here, $z$ is the coordinate of an emergent extra-dimensional space, which corresponds to the iteration step of the RG transformations. $G_{xx'}[g_{\mu\nu}(x,z),\varphi(x,z)]$ is the Green's function of high-energy quantum fluctuations of $\phi_{\alpha}(x)$, given by
\bqa && \Big\{- \frac{1}{\sqrt{g(x,z)}} \partial_{\mu} \Big( \sqrt{g(x,z)} g^{\mu\nu}(x,z) \partial_{\nu} \Big) + \frac{1}{\epsilon} [m^{2} - i \varphi(x,z)] \Big\} G_{xx'}[g_{\mu\nu}(x,z),\varphi(x,z)] = \frac{1}{\sqrt{g(x,z)}} \delta^{(D)}(x-x') . \eqa
$\epsilon$ is an energy scale for the RG transformation proportional to $d z$, which controls quantum fluctuations of high-energy modes of original matter fields. It is this Green's function that encodes the dynamical information in the recursive RG transformation.

The recursive RG transformation terminates at $z = z_{f}$ and the resulting renormalized metric tensor appears in the effective action $\mathcal{S}_{IR}[\varphi(x,z_{f})] = \int d^{D} x \sqrt{g(x,z_{f})} \Big\{ g^{\mu\nu}(x,z_{f}) [\partial_{\mu} \phi_{\alpha}(x)] [\partial_{\nu} \phi_{\alpha}(x)] + [m^{2} - i \varphi(x,z_{f})] \phi_{\alpha}^{2}(x) \Big\}$. Here, $z_{f}$ is a phenomenologically introduced IR cut-off energy scale, given by $z_{f} = f \epsilon$, where $f$ is the iteration number of recursive RG transformations. Although we can stop these recursive RG transformations at any energy scale $z_{f}$, the full integration is achieved by $z_{f} = \Lambda_{UV}$, where $\Lambda_{UV}$ is the UV cut-off energy scale. For example, suppose that $\Lambda_{UV}$ is a UV cut-off scale for dynamically fluctuating field variables in momentum space. Then, $z_{f} = \Lambda_{UV}$ covers the whole region of the momentum space for such quantum fields, where the full integration up to $z_{f} = \Lambda_{UV}$ completes the recursive RG transformations. In this respect $z_{f}$ may be considered as the UV cut-off scale rather than the IR cut-off scale although physically it corresponds to the IR scale. The renormalized metric tensor at $z = z_{f}$ describes not only the wave-function renormalization but also the mass renormalization effects \cite{Holographic_Description_RG_GR}, to be clarified in section IV. Here, we point out our gauge fixing choice $g^{DD}(x,z) = 1$ and $g^{\mu D}(x,z) = g^{D \nu}(x,z) = 0$, where $\mu$ and $\nu$ run from $0$ to $D - 1$. Performing the path integral with respect to the original matter field $\phi_{\alpha}(x)$, we obtain
\bqa && \mathcal{S}_{IR}[\varphi(x,z_{f})] = - \frac{N}{2} \mbox{tr}_{xx'} \ln \sqrt{g(x,z_{f})} \Big\{- \frac{1}{\sqrt{g(x,z_{f})}} \partial_{\mu} \Big( \sqrt{g(x,z_{f})} g^{\mu\nu}(x,z_{f}) \partial_{\nu} \Big) + [m^{2} - i \varphi(x,z_{f})] \Big\} \nn && \approx N \int d^{D} x \sqrt{g(x,z_{f})} \Big\{ \frac{\mathcal{C}_{\varphi}^{f}}{2} g^{\mu\nu}(x,z_{f}) [\partial_{\mu} \varphi(x,z_{f})] [\partial_{\nu} \varphi(x,z_{f})] + \frac{1}{2 \kappa_{f}} \Big( R(x,z_{f}) - 2 \Lambda_{f} \Big) \Big\} , \nonumber \eqa
where the gradient expansion with respect to $m$ has been performed. $\mathcal{C}_{\varphi}^{f}$ is a positive numerical coefficient, which decreases as $m$ increases. The last Einstein-Hilbert action results from vacuum fluctuations of quantum matter fields, referred to as induced gravity \cite{Gradient_Expansion_Gravity_I,Gradient_Expansion_Gravity_II}.

The bulk effective action originates from quantum fluctuations of the high-energy modes of $\phi_{\alpha}(x)$ in the recursive RG transformations. Actually, the induced gravity action and the kinetic energy of the dual scalar field at a given slice $z$ of the bulk result from
\bqa && \mathcal{S}_{Bulk}[\varphi(x,z);z] = - \frac{N}{2} \mbox{tr}_{xx'} \ln \sqrt{g(x,z)} \Big\{- \frac{1}{\sqrt{g(x,z)}} \partial_{\mu} \Big( \sqrt{g(x,z)} g^{\mu\nu}(x,z) \partial_{\nu} \Big) + \frac{1}{\epsilon} [m^{2} - i \varphi(x,z)] \Big\} \nn && \approx \epsilon N \int d^{D} x \sqrt{g(x,z)} \Big\{ \frac{\mathcal{C}_{\varphi}}{2} g^{\mu\nu}(x,z) [\partial_{\mu} \varphi(x,z)] [\partial_{\nu} \varphi(x,z)] + \frac{1}{2 \kappa} \Big( R(x,z) - 2 \Lambda \Big) \Big\} , \nonumber \eqa
where $\epsilon$ is the energy scale of the RG transformation as pointed out before. Here, $\epsilon \sum_{z = 1}^{f}$ may be replaced with $\int_{0}^{z_{f}} d z$ in the continuum representation for the extra-dimensional coordinate $z$. As a result, the bulk effective action is given by
\bqa && \mathcal{S}_{Bulk}[\varphi(x,z)] = N \int_{0}^{z_{f}} d z \int d^{D} x \sqrt{g(x,z)} \Big\{ \frac{1}{2u} [\partial_{z} \varphi(x,z)]^{2} + \frac{\mathcal{C}_{\varphi}}{2} g^{\mu\nu}(x,z) [\partial_{\mu} \varphi(x,z)] [\partial_{\nu} \varphi(x,z)] + \frac{1}{2 \kappa} \Big( R(x,z) - 2 \Lambda \Big) \Big\} . \nonumber \eqa
Since we would like to emphasize the physical meaning of the IR boundary condition in section III.B, we kept the superscript $f$ for both the induced gravity action and the kinetic energy term of the dual scalar field at the IR boundary. However, all these coefficients are the same as those of the bulk effective action. As a result, the bulk effective action is smoothly connected with the IR boundary action with the gauge fixing mentioned above.

This effective action plays a central role in the dynamics of the collective dual field variable $\varphi(x,z)$. Taking the large $N$ limit, quantum fluctuations of these scalar bosons are suppressed. Performing the variation of this effective action with respect to $\varphi(x,z)$, $\partial_{z} \varphi(x,z)$, and $\partial_{\mu} \varphi(x,z)$, we obtain a Lagrange equation of motion for the collective dual field variable. As discussed in the introduction, this equation of motion describes the RG flow of the order parameter field, thus identified with the Callan-Symanzik equation of a one-particle Green's function of the original matter field. Since it is the second-order differential equation, we need two boundary conditions. The UV boundary condition is given by the variation of the effective action in Eq. (\ref{Partition_Function_Metric_RG_Flow}) with respect to $\varphi(x,0)$ and $[\partial_{z} \varphi(x,z)]_{z = 0}$. In the same way the IR boundary condition is given by the variation of the effective action in Eq. (\ref{Partition_Function_Metric_RG_Flow}) with respect to $\varphi(x,z_{f})$ and $[\partial_{z} \varphi(x,z)]_{z = z_{f}}$. It turns out that the IR boundary condition is reduced into the conventional mean-field equation for the order parameter field in the limit of $z_{f} \rightarrow 0$, where renormalization effects are not introduced. In other words, the IR boundary condition takes into account not only the renormalization effects of interaction vertices but also the RG flow of the order parameter field in a self-consistent intertwined way.

It is necessary to discuss the physical origin of the emergent cosmological constant $\Lambda$. It may be regarded as a vacuum-fluctuation energy, here given by quantum fluctuations of high-energy degrees of freedom, which are accumulated in the recursive RG transformation. More precisely, high-energy fluctuations of original matter fields $\phi_{\alpha}(x)$ give rise to the following effective action at a fixed slice $z$ in the continuum-coordinate representation of $z$
\bqa && - \frac{N}{2} \mbox{tr}_{xx'} \ln \sqrt{g(x,z)} \Big\{- \frac{1}{\sqrt{g(x,z)}} \partial_{\mu} \Big( \sqrt{g(x,z)} g^{\mu\nu}(x,z) \partial_{\nu} \Big) + \frac{1}{\epsilon} [m^{2} - i \varphi(x,z)] \Big\} , \nonumber \eqa
as discussed above. We pointed out that both the Einstein-Hilbert action and the kinetic energy of the dual scalar field result from the gradient expansion of this effective action \cite{Gradient_Expansion_Gravity_I,Gradient_Expansion_Gravity_II}. As a result, the bare cosmological constant is given by
\bqa && \frac{\Lambda}{\kappa} = - \frac{1}{2} \ln m^{2} . \eqa
We point out that this is the result of the large $N$ limit: Although there appear several terms more, which come from RG transformations for other dynamical field variables, such terms are less dominant in the large $N$ limit, and thus neglected. This expression indicates that the sign of the bare cosmological constant changes, approaching the quantum critical point. In our sign convention, $\Lambda > 0$ near the quantum critical point corresponds to a negative cosmological constant. However, it is also important to check out how this cosmological constant renormalizes in the quantum gravity theory beyond the large $N$ limit. For the case of fermions, the overall sign becomes opposite at least for this mass contribution \cite{Holographic_Description_Einstein_Maxwell}.

One may point out that the metric tensor is not dynamical, where the RG flow of the metric tensor is given by the $\delta-$function constraint. In this respect the Einstein-Hilbert action has nothing to do with the dynamics of the emergent metric tensor, which serves as a vacuum energy only. The absence of the metric-tensor dynamics results from the fact that effective interactions between energy-momentum tensor currents are not taken into account from the start here. We recall that the bulk dynamics of the dual scalar field $\varphi(x,z)$ given by the kinetic energy originates from the recursive RG transformation with the Hubbard-Stratonovich transformation for the effective interaction term of the density-density ($O(N)$ singlet) channel. In this respect we have to introduce an effective interaction term between energy-momentum tensor currents into the effective action from the start in order to promote the metric tensor to be dynamical. Actually, this point has been clearly discussed in Ref. \cite{Holographic_Description_Einstein_Klein_Gordon}, following the proposal of Ref. \cite{Dynamical_Metric}. As a result of the introduction of this tensor-type effective interaction, we obtain the kinetic energy of the metric tensor, which results in the equation of motion for the emergent metric tensor from the minimization of the free-energy functional with respect to the metric tensor, given by the second-order differential equation with respect to the emergent extra-dimensional space. In this case the effective Einstein-Hilbert action gives rise to an Einstein equation at a fixed slice along the extra-dimensional space.

Solving both the RG flow of the metric tensor with the Green's function and the equation of motion of the dual scalar field with UV and IR boundary conditions, we can reveal non-perturbative physics not only near but also away from the quantum critical point of this effective field theory. As a result, one may have interesting insight from geometrical interpretation of a quantum phase transition, where a ``horizon" solution appears to describe the quantum phase transition \cite{Holographic_Description_Entanglement_Entropy}. This theoretical framework may shed light on even the problem of the black hole entropy \cite{Entanglement_Entropy_SungSik}. However, it is not convenient to consider a field theory from the first for the application to the condensed matter physics. It is necessary to start from a lattice model. Here, we consider a one-dimensional lattice model
\bqa && Z = \int D \Phi_{i}(\tau) \exp\Big[ - \int_{0}^{\beta} d \tau \sum_{i = 1}^{L} \Big\{ \Big(\partial_{\tau} \Phi_{i}(\tau) \Big)^{2} - t \Big( \Phi_{i}(\tau) \Phi_{i+1}(\tau) + \Phi_{i+1}(\tau) \Phi_{i}(\tau) \Big) + m^{2} [\Phi_{i}(\tau)]^{2} + \frac{u}{2} [\Phi_{i}(\tau)]^{4} \Big\} \Big] . \nn \label{Partition_Function_Coupling} \eqa
$\Phi_{i}(\tau)$ is a real scalar field at site $i$. $t$ is the hopping integral between nearest neighbor sites. $m$ is the mass of this scalar field and $u$ is the strength of self-interactions between these scalar bosons. This model may be regarded as an effective lattice field theory of the transverse-field Ising model \cite{Kadanoff_RG}.

To obtain an effective holographic dual field theory, we apply the Kadanoff block-spin transformation \cite{Kadanoff_RG} in a recursive way. First, we perform the Hubbard-Stratonovich transformation for the self-interaction term and reformulate it in terms of a collective dual field variable. Second, we separate all dynamical field variables into those on even and odd sites. Third, we perform the path integral with respect to all dynamical fields on even sites and obtain an effective lattice field theory in terms of the original matter field and the collective dual field variable on odd sites, where their effective interactions are renormalized by quantum fluctuations of even-site fields. Since the path integral with respect to the order parameter field gives rise to effective self-interactions between the original matter fields, fourth, we perform the Hubbard-Stratonovich transformation for these effective interactions once again and rewrite this term in terms of the second collective dual field variable. Fifth, we update all the coupling constants with rescaling to return the effective action to its original form. As a result, we find a recursive expression of the effective action in the partition function, which consists of the RG $\beta-$functions of the coupling constants and the RG flow of the order parameter fields. Taking the large $N$ limit when the flavor degeneracy is introduced into the above lattice field theory, the coupled equations of motion for both order parameter fields describe the RG flow of the order parameter field, identified with the Callan-Symanzik equation for the order parameter field. This mean-field theoretical framework is beyond the conventional mean-field theory in the respect that renormalization effects of both effective interactions and order parameter fields are taken into account in a self-consistent intertwined way.

Repeating these recursive RG transformations a la Kadanoff and reformulating the iteration steps of the RG transformations with a continuous coordinate $z$, we obtain an effective holographic dual field theory as follows
\bqa && Z = \int D \Phi_{i}(\tau) D \varphi(z) D m^{2}(z) D t(z) D u(z) \delta \Big(m^{2}(0) - m^{2}\Big) \delta \Big(t(0) - t\Big) \delta \Big(u(0) - u\Big) \nn && \delta \Bigg( \partial_{z} m^{2}(z) + \frac{2 m^{2}(z) [t(z)]^{2}}{a \Big(2 [m^{2}(z)]^{2} + u(z)\Big) } \Bigg) \delta \Bigg( \partial_{z} t(z) + \frac{1}{a} t(z) - \frac{m^{2}(z) [t(z)]^{2}}{a\Big(2 [m^{2}(z)]^{2} + u(z)\Big)} - \frac{i}{2} \partial_{z} \varphi(z) \Bigg) \nn && \delta \Bigg( \partial_{z} u(z) + \frac{1}{a} u(z) - \frac{u(z) [t(z)]^{4}}{2 a [m^{2}(z)]^{2} \Big(2 [m^{2}(z)]^{2} + u(z)\Big)} \Bigg) \nn && \exp\Big[ - \int_{0}^{\beta} d \tau \sum_{i = 1}^{L} \Big\{ \Big(\partial_{\tau} \Phi_{i}(\tau)\Big)^{2} + \Big( m^{2}(z_{f}) - i \varphi(z_{f}) \Big) [\Phi_{i}(\tau)]^{2} - t(z_{f}) \Big( \Phi_{i}(\tau) \Phi_{i+1}(\tau) + \Phi_{i+1}(\tau) \Phi_{i}(\tau) \Big) \Big\} \nn && - \int_{0}^{\beta} d \tau \sum_{i = 1}^{L} \frac{1}{2u(0)} [\varphi(0)]^{2} \nn && - \int_{0}^{z_{f}} d z \int_{0}^{\beta} d \tau \sum_{i = 1}^{L} \Big\{ \frac{a}{2 u(z)} \Big( \partial_{z} \varphi(z)\Big)^{2} + \frac{u(z)}{2 a \Big(2 [m^{2}(z)]^{2} + u(z)\Big)} + \frac{1}{2 a \beta} \ln \Big\{ 2 \sinh \Big( \frac{\beta}{2} m(z) \Big) \Big\} \Big\} \Big] . \label{Partition_Function_Coupling_RG_Flow} \eqa
We emphasize that this effective partition function is completely identical to the original one Eq. (\ref{Partition_Function_Coupling}) up to a constant value. First three $\delta-$function constraints indicate three initial conditions for the coupling constants of mass, nearest-neighbor hopping integral, and self-interaction, respectively. Next three $\delta-$function constraints impose the RG flow equations of the coupling constants, given by the first-order differential equation as expected. These RG flow equations terminate at $z = z_{f}$ and fully renormalized values of the coupling constants appear in the IR effective action $\mathcal{S}_{IR}[\varphi(z_{f})] = \int_{0}^{\beta} d \tau \sum_{i = 1}^{L} \Big\{ \Big(\partial_{\tau} \Phi_{i}(\tau)\Big)^{2} + \Big( m^{2}(z_{f}) - i \varphi(z_{f}) \Big) [\Phi_{i}(\tau)]^{2} - t(z_{f}) \Big( \Phi_{i}(\tau) \Phi_{i+1}(\tau) + \Phi_{i+1}(\tau) \Phi_{i}(\tau) \Big) \Big\}$.

Neglecting quantum fluctuations of the collective dual field variables, or more formally, taking the large $N$ limit when the flavor degeneracy is introduced into the original matter field, we obtain a classical equation of motion for the order parameter field, given by the variation of the bulk effective action $\mathcal{S}_{Bulk}[\varphi(z)] = \int_{0}^{z_{f}} d z \int_{0}^{\beta} d \tau \sum_{i = 1}^{L} \Big\{ \frac{a}{2 u(z)} \Big( \partial_{z} \varphi(z)\Big)^{2} + \frac{u(z)}{2 a \Big(2 [m^{2}(z)]^{2} + u(z)\Big)} + \frac{1}{2 a \beta} \ln \Big\{ 2 \sinh \Big( \frac{\beta}{2} m(z) \Big) \Big\} \Big\}$ with respect to  $\varphi(z)$ and $\partial_{z} \varphi(z)$. It is the second-order differential equation with respect to the emergent extra-dimensional space. Both UV and IR boundary conditions are given by the variation of the effective action with respect to $\varphi(0)$ and $[\partial_{z} \varphi(z)]_{z = 0}$ for UV and $\varphi(z_{f})$ and $[\partial_{z} \varphi(z)]_{z = z_{f}}$ for IR. As will be shown in the next section, the IR boundary condition generalizes the conventional mean-field equation in a non-perturbative way, introducing not only renormalization effects of the coupling constants but also the RG flow of the order parameter field in an intertwined way.

In this paper we show the equivalence between Eq. (\ref{Partition_Function_Metric_RG_Flow}) and Eq. (\ref{Partition_Function_Coupling_RG_Flow}) up to a normalization constant. Since Eq. (\ref{Partition_Function_Metric}) is identical to Eq. (\ref{Partition_Function_Coupling}) up to a constant, these two dual holographic partition functions have to be the same. In section \ref{Holography_Lattice}, we show derivation from Eq. (\ref{Partition_Function_Coupling}) to Eq. (\ref{Partition_Function_Coupling_RG_Flow}). In section \ref{Equivalence_Lattice_Continuum}, we prove the equivalence between Eq. (\ref{Partition_Function_Metric_RG_Flow}) and Eq. (\ref{Partition_Function_Coupling_RG_Flow}). In section \ref{QCP_AdS3}, we show the emergence of AdS$_{3}$ geometry at the quantum critical point of the transverse-field Ising model. In section \ref{Discussion}, we consider the Majorana-fermion representation of the transverse-field Ising model \cite{Kitaev_Model} and discuss the boson-fermion duality in a geometrical point of view.

%
%

%
%
%
%
%

%
%

%
%
%
%

\section{A dual holographic effective field theory for a one-dimensional lattice field theory of interacting bosons} \label{Holography_Lattice}

\subsection{Derivation of an effective lattice field theory Eq. (\ref{Partition_Function_Coupling_RG_Flow}) from Eq. (\ref{Partition_Function_Coupling})}

We perform the Hubbard-Stratonovich transformation for the self-interaction term in Eq. (\ref{Partition_Function_Coupling}). Then, we obtain
\bqa && Z = \int D \Phi_{i}(\tau) D \varphi^{(0)} \exp\Big[ - \int_{0}^{\beta} d \tau \sum_{i = 1}^{L} \Big\{ \Big(\partial_{\tau} \Phi_{i}(\tau)\Big)^{2} + \Big( m^{2} - i \varphi^{(0)} \Big) [\Phi_{i}(\tau)]^{2} - t \Big( \Phi_{i}(\tau) \Phi_{i+1}(\tau) + \Phi_{i+1}(\tau) \Phi_{i}(\tau) \Big) \nn && + \frac{1}{2u} \varphi^{(0) 2} \Big\} \Big] , \eqa
where $\varphi^{(0)}$ is the collective dual field variable.

To implement the Kadanoff block-spin transformation, we separate all dynamical fields into those on even and odd sites. Then, we perform the path integral with respect to quantum fluctuations on even sites. Performing the path integral for even-site original matter fields, we obtain the vacuum-fluctuation contribution of even-site boson fields given by $\frac{1}{2} \mbox{tr}_{\tau\tau'} \ln \Big( - \partial_{\tau}^{2} + m^{2} - i \varphi_{2l}^{(0)} \Big)$ and an effective coupling term between next-nearest-neighbor-site bosons given by $- \frac{t^{2}}{2} \Big( \Phi_{2l-1} + \Phi_{2l+1} \Big) \Big( - \partial_{\tau}^{2} + m^{2} - i \varphi_{2l}^{(0)} \Big)^{-1} \Big( \Phi_{2l-1} + \Phi_{2l+1} \Big)$. Here, $2l$ and $2l-1$ represent even- and odd-site, respectively. Performing the gradient expansion for these two effective actions with respect to $i \varphi_{2l}^{(0)} / m^{2}$ up to the second order, and doing the path integral for the collective dual scalar field $\varphi_{2l}^{(0)}$ in the Gaussian level, we obtain
\bqa && Z = \int D \Phi_{i}(\tau) D \varphi^{(0)} \exp\Big[ - \int_{0}^{\beta} d \tau \sum_{i = 1}^{L} \Big\{ \Big(\partial_{\tau} \Phi_{i}(\tau)\Big)^{2} + \Big( m^{2} - \frac{2 m^{2} t^{2}}{ 2 [m^{2}]^{2} + u } - i \varphi^{(0)} \Big) [\Phi_{i}(\tau)]^{2} \nn && - \frac{m^{2} t^{2}}{ 2 [m^{2}]^{2} + u } \Big( \Phi_{i}(\tau) \Phi_{i+1}(\tau) + \Phi_{i+1}(\tau) \Phi_{i}(\tau) \Big) + \frac{1}{2u} \varphi^{(0) 2} \nn && + \frac{u t^{4}}{16 [m^{2}]^{2} \Big(2 [m^{2}]^{2} + u\Big)} \Big( 2 [\Phi_{i}(\tau)]^{2} + \Phi_{i}(\tau) \Phi_{i+1}(\tau) + \Phi_{i+1}(\tau) \Phi_{i}(\tau) \Big)^{2} \nn && + \frac{u}{2 \Big(2 [m^{2}]^{2} + u\Big)} + \frac{1}{4 \beta} \Big\{ \ln \Big( 1 - e^{ - \beta m} \Big) + \ln \Big( e^{\beta m} - 1 \Big) \Big\} \Big\} \Big] , \eqa
where rescaling has been performed. This RG transformation procedure is straightforward and justified as far as $i \varphi_{2l}^{(0)} / m^{2} \ll 1$ is satisfied. Quantum fluctuations of even-site original matter fields give rise to renormalization effects on all the coupling constants of the mass parameter, the hopping integral, and the self-interaction parameter. Even-site order-parameter fluctuations result in effective interactions for the original matter fields, given by $\mathcal{V}_{eff} = \frac{u t^{4}}{16 [m^{2}]^{2} \Big(2 [m^{2}]^{2} + u\Big)} \Big( 2 [\Phi_{i}(\tau)]^{2} + \Phi_{i}(\tau) \Phi_{i+1}(\tau) + \Phi_{i+1}(\tau) \Phi_{i}(\tau) \Big)^{2}$.

To return this expression into the original form of the effective action, we perform the Hubbard-Stratonovich transformation once again for this newly generated effective interaction. As a result, we obtain the following expression
\bqa && Z = \int D \Phi_{i}(\tau) D \varphi^{(1)} D \varphi^{(0)} \exp\Big[ - \int_{0}^{\beta} d \tau \sum_{i = 1}^{L} \Big\{ \Big(\partial_{\tau} \Phi_{i}(\tau)\Big)^{2} + \Big( m^{2} - \frac{2 m^{2} t^{2}}{ 2 [m^{2}]^{2} + u } - i \varphi^{(1)} \Big) [\Phi_{i}(\tau)]^{2} \nn && - \Big\{ \frac{m^{2} t^{2}}{ 2 [m^{2}]^{2} + u } + \frac{i}{2} \Big( \varphi^{(1)} - \varphi^{(0)} \Big) \Big\} \Big( \Phi_{i}(\tau) \Phi_{i+1}(\tau) + \Phi_{i+1}(\tau) \Phi_{i}(\tau) \Big) + \frac{1}{2u} \varphi^{(0) 2} + \frac{[m^{2}]^{2} \Big(2 [m^{2}]^{2} + u\Big)}{u t^{4}} \Big( \varphi^{(1)} - \varphi^{(0)} \Big)^{2} \nn && + \frac{u}{2 \Big(2 [m^{2}]^{2} + u\Big)} + \frac{1}{4 \beta} \Big\{ \ln \Big( 1 - e^{ - \beta m} \Big) + \ln \Big( e^{\beta m} - 1 \Big) \Big\} \Big\} \Big] , \eqa
where the second dual field variable $\varphi^{(1)}$ has been introduced.

We update all the coupling constants as follows
\bqa && Z = \int D \Phi_{i}(\tau) D \varphi^{(1)} D \varphi^{(0)} D m^{2 (1)} D m^{2 (0)} D t^{(1)} D t^{(0)} D u^{(1)} D u^{(0)} \delta \Big(m^{2 (0)} - m^{2}\Big) \delta \Big(t^{(0)} - t\Big) \delta \Big(u^{(0)} - u\Big) \nn && \delta \Bigg( m^{2 (1)} - m^{2 (0)} + \frac{2 m^{2 (0)} [t^{(0)}]^{2}}{ 2 [m^{2 (0)}]^{2} + u^{(0)} } \Bigg) \delta \Bigg( t^{(1)} - \frac{m^{2 (0)} [t^{(0)}]^{2}}{ 2 [m^{2 (0)}]^{2} + u^{(0)} } - \frac{i}{2} \Big( \varphi^{(1)} - \varphi^{(0)} \Big) \Bigg) \nn && \delta \Bigg( u^{(1)} - \frac{u^{(0)} [t^{(0)}]^{4}}{2 [m^{2 (0)}]^{2} \Big(2 [m^{2 (0)}]^{2} + u^{(0)}\Big)} \Bigg) \nn && \exp\Big[ - \int_{0}^{\beta} d \tau \sum_{i = 1}^{L} \Big\{ \Big(\partial_{\tau} \Phi_{i}(\tau)\Big)^{2} + \Big( m^{2 (1)} - i \varphi^{(1)} \Big) [\Phi_{i}(\tau)]^{2} - t^{(1)} \Big( \Phi_{i}(\tau) \Phi_{i+1}(\tau) + \Phi_{i+1}(\tau) \Phi_{i}(\tau) \Big) \Big\} \nn && - \int_{0}^{\beta} d \tau \sum_{i = 1}^{L} \frac{1}{2u^{(0)}} [\varphi^{(0)}]^{2} \nn && - \int_{0}^{\beta} d \tau \sum_{i = 1}^{L} \Big\{ \frac{1}{2 u^{(1)}} \Big( \varphi^{(1)} - \varphi^{(0)} \Big)^{2} + \frac{u^{(0)}}{2 \Big(2 [m^{2 (0)}]^{2} + u^{(0)}\Big)} + \frac{1}{2 \beta} \ln \Big\{ 2 \sinh \Big( \frac{\beta}{2} m^{(0)} \Big) \Big\} \Big\} \Big] , \eqa
where the first three $\delta-$function constraints denote the initial values of the coupling constants and the last three $\delta-$function constraints impose their renormalization equations. We point out that the resulting effective action is now written in terms of these updated coupling constants, essentially the same form as the original action.

We repeat these RG transformations. As a result, we obtain the following recursive expression for the effective action in the partition function
\bqa && Z = \int D \Phi_{i}(\tau) D \varphi^{(0)} D m^{2 (0)} D t^{(0)} D u^{(0)} \delta \Big(m^{2 (0)} - m^{2}\Big) \delta \Big(t^{(0)} - t\Big) \delta \Big(u^{(0)} - u\Big) \nn && \int \Pi_{k = 1}^{f} D \varphi^{(k)} D m^{2 (k)} D t^{(k)} D u^{(k)} \delta \Bigg( \frac{m^{2 (k)} - m^{2 (k-1)}}{a} + \frac{2 m^{2 (k-1)} [t^{(k-1)}]^{2}}{a \Big(2 [m^{2 (k-1)}]^{2} + u^{(k-1)}\Big) } \Bigg) \nn && \delta \Bigg( \frac{t^{(k)} - t^{(k-1)}}{a} + \frac{1}{a} t^{(k-1)} - \frac{m^{2 (k-1)} [t^{(k-1)}]^{2}}{a\Big(2 [m^{2 (k-1)}]^{2} + u^{(k-1)}\Big)} - \frac{i}{2} \frac{ \varphi^{(k)} - \varphi^{(k-1)} }{a} \Bigg) \nn && \delta \Bigg( \frac{u^{(k)} - u^{(k-1)}}{a} + \frac{1}{a} u^{(k-1)} - \frac{u^{(k-1)} [t^{(k-1)}]^{4}}{2 a [m^{2 (k-1)}]^{2} \Big(2 [m^{2 (k-1)}]^{2} + u^{(k-1)}\Big)} \Bigg) \nn && \exp\Big[ - \int_{0}^{\beta} d \tau \sum_{i = 1}^{L} \Big\{ \Big(\partial_{\tau} \Phi_{i}(\tau)\Big)^{2} + \Big( m^{2 (f)} - i \varphi^{(f)} \Big) [\Phi_{i}(\tau)]^{2} - t^{(f)} \Big( \Phi_{i}(\tau) \Phi_{i+1}(\tau) + \Phi_{i+1}(\tau) \Phi_{i}(\tau) \Big) \Big\} \nn && - \int_{0}^{\beta} d \tau \sum_{i = 1}^{L} \frac{1}{2u^{(0)}} [\varphi^{(0)}]^{2} \nn && - a \sum_{k = 1}^{f} \int_{0}^{\beta} d \tau \sum_{i = 1}^{L} \Big\{ \frac{a}{2 u^{(k)}} \Big( \frac{\varphi^{(k)} - \varphi^{(k-1)}}{a} \Big)^{2} + \frac{u^{(k-1)}}{2 a \Big(2 [m^{2 (k-1)}]^{2} + u^{(k-1)}\Big)} + \frac{1}{2 a \beta} \ln \Big\{ 2 \sinh \Big( \frac{\beta}{2} m^{(k-1)} \Big) \Big\} \Big\} \Big] . \eqa
Here, $a$ is introduced to play the role of an energy scale in the RG transformation, to be clarified below.

The final step is to reformulate this recursive expression as the form of a differential equation. It is conventional to replace $a \sum_{k = 1}^{f}$ with $\int_{0}^{z_{f}} d z$. In addition, we change $\frac{\varphi^{(k)} - \varphi^{(k-1)}}{a}$ to $\partial_{z} \varphi(z)$, where the iteration step $(k)$ of the RG transformation appears as an extra-dimensional space $z$ with an energy scale $a$. As a result, we obtain Eq. (\ref{Partition_Function_Coupling_RG_Flow}) as an effective dual holographic theory.

One may point out that the dual scalar order-parameter field $\varphi^{(k)}$ has to be a spacetime dependent function. Although this criticism does make sense in principle, here we did not introduce it intentionally. More precisely speaking, as far as translational symmetry is preserved, we are allowed to focus on such spacetime independent solutions for the dual scalar field. Actually, such solutions are realized as far as we concern translationally invariant vacuum solutions. However, if we consider a spacetime dependent external field $\varphi_{i}^{ext}(\tau)$, we have to introduce the spacetime dependence of such dynamically fluctuating fields more explicitly. In this case, we also have to introduce the spacetime dependence in all the coupling constants such as the renormalized hopping integral, the renormalized mass parameter, and the renormalized effective interaction parameter. Then, it becomes more complex to compare the path-integral expressions of the two partition functions. In this resect we focus on the case of translational invariance for comparison.

\subsection{Remarks on this dual holographic effective field theory}

In this paper we do not perform detail analysis on this problem. Instead, we focus on how the RG flow is translated into geometry. However, it is necessary to discuss how this theoretical framework describes non-perturbative physics of the quantum phase transition in a mean-field fashion.

Performing the path integral with respect to the original matter fields, we obtain an effective mean-field free-energy functional
\bqa && \frac{1}{L} F[\varphi(z)] = \frac{1}{\beta} \frac{1}{L} \sum_{k} \ln \Bigg\{ 2 \sinh \Bigg( \frac{\beta}{2} \sqrt{- 2 t(z_{f}) \cos k + m^{2}(z_{f}) - i \varphi(z_{f}) } \Bigg) \Bigg\} + \frac{1}{2u(0)} [\varphi(0)]^{2} \nn && + \frac{1}{a} \int_{0}^{z_{f}} d z \Bigg( \frac{a^{2}}{2 u(z)} \Big( \partial_{z} \varphi(z)\Big)^{2} + \frac{u(z)}{2 \Big(2 [m^{2}(z)]^{2} + u(z)\Big)} + \frac{1}{2 \beta} \ln \Big\{ 2 \sinh \Big( \frac{\beta}{2} m(z) \Big) \Big\} \Bigg) . \eqa
The first term is a typical form of the free energy for relativistic massive scalar fields, where their velocity and effective mass are renormalized and described by the RG flow equations
\bqa && a \partial_{z} m^{2}(z) = - \frac{2 m^{2}(z) [t(z)]^{2}}{2 [m^{2}(z)]^{2} + u(z)} \eqa
for the mass parameter and
\bqa && a \partial_{z} t(z) = - t(z) + \frac{m^{2}(z) [t(z)]^{2}}{ 2 [m^{2}(z)]^{2} + u(z) } + \frac{i}{2} a \partial_{z} \varphi(z) \eqa
for the hopping integral. The RG flow of the self-interaction coefficient appears in the bulk sector of the free energy, given by
\bqa && a \partial_{z} u(z) = - u(z) + \frac{u(z) [t(z)]^{4}}{2 [m^{2}(z)]^{2} \Big(2 [m^{2}(z)]^{2} + u(z)\Big)} , \eqa
which affects the RG flow of the order parameter field $\varphi(z)$.

%
%

The last equation is the Lagrange equation of motion for the order parameter field, given by the variation of this free-energy functional with respect to $\varphi(z)$ and $\partial_{z} \varphi(z)$. Then, we obtain
\bqa && - \partial_{z}^{2} \varphi(z) + [\partial_{z} \ln u(z)] [\partial_{z} \varphi(z)] = 0 . \eqa
%
%
%
It is straightforward to solve this differential equation, the solution of which is
\bqa && \varphi(z) = \varphi(0) + \frac{[\partial_{z} \varphi(z)]_{z = 0}}{u(0)} \int_{0}^{z} d z' u(z') . \eqa
Here, there are two unknown coefficients $\varphi(0)$ and $[\partial_{z} \varphi(z)]_{z = 0}$ as expected. These constants are determined by two boundary conditions, given by the self-consistency of this theoretical framework.

The first boundary condition is fixed by the first RG transformation, given by the so called UV free energy
%
%
\bqa && F_{UV} = \frac{1}{2u(0)} [\varphi(0)]^{2} - \frac{a}{2 u(0)} \varphi(0) [\partial_{z} \varphi(z)]_{z = 0} . \eqa
Here, the second term results from the UV boundary contribution of the bulk effective action. Performing the variation of this boundary free energy with respect to $\varphi(0)$ and $[\partial_{z} \varphi(z)]_{z = 0}$, we obtain the corresponding UV boundary condition, which gives $\varphi(0) = 0$. The resulting solution is
\bqa && \varphi(z) = \frac{[\partial_{z} \varphi(z)]_{z = 0}}{u(0)} \frac{1}{a} \int_{0}^{z} d z' u(z') . \eqa

%
%
%
%

The second boundary condition is fixed by the self-consistency of the last RG transformation in this effective field theory, given by the so called IR free energy
\bqa && F_{IR} = \frac{1}{\beta} \frac{1}{L} \sum_{k} \ln \Bigg\{ 2 \sinh \Bigg( \frac{\beta}{2} \sqrt{- 2 t(z_{f}) \cos k + m^{2}(z_{f}) - i \varphi(z_{f}) } \Bigg) \Bigg\} - \frac{a}{2 u(z_{f})} [i \varphi(z_{f})] [i \partial_{z} \varphi(z)]_{z = z_{f}} . \eqa
Here, the second term results from the IR boundary contribution of the bulk effective action. Performing the variation of this boundary free energy with respect to $\varphi(z_{f})$ and $[\partial_{z} \varphi(z)]_{z = z_{f}}$, we obtain the corresponding IR boundary condition
\bqa && - \frac{1}{4} \frac{1}{L} \sum_{k} \frac{1}{\sqrt{- 2 t(z_{f}) \cos k + m^{2}(z_{f}) - i \varphi(z_{f}) }} \coth \Bigg( \frac{\beta}{2} \sqrt{- 2 t(z_{f}) \cos k + m^{2}(z_{f}) - i \varphi(z_{f}) } \Bigg) = \frac{[a \partial_{z} u(z)]_{z = z_{f}}}{2 [u(z_{f})]^{2}} [i \varphi(z_{f})] . \nn \eqa

This IR boundary condition reminds us of the conventional mean-field equation. To verify how this expression generalizes the conventional mean-field equation of the order parameter field, we replace $[a \partial_{z} u(z)]_{z = z_{f}}$ with the RG flow of the self-interaction parameter
%
%
%
\bqa && [a \partial_{z} u(z)]_{z = z_{f}} = - u(z_{f}) + \frac{u(z_{f}) [t(z_{f})]^{4}}{2 [m^{2}(z_{f})]^{2} \Big(2 [m^{2}(z_{f})]^{2} + u(z_{f})\Big)} \eqa
in this IR boundary condition. As a result, we obtain
\bqa && \frac{1}{4} \frac{1}{L} \sum_{k} \frac{1}{\sqrt{- 2 t(z_{f}) \cos k + m^{2}(z_{f}) - i \varphi(z_{f}) }} \coth \Bigg( \frac{\beta}{2} \sqrt{- 2 t(z_{f}) \cos k + m^{2}(z_{f}) - i \varphi(z_{f}) } \Bigg) \nn && = \frac{[i \varphi(z_{f})]}{2 u(z_{f})} - \frac{[t(z_{f})]^{4}}{4 u(z_{f}) [m^{2}(z_{f})]^{2} \Big(2 [m^{2}(z_{f})]^{2} + u(z_{f})\Big)} [i \varphi(z_{f})] . \label{IR_Boundary_Condition_Nonperturbativeness} \eqa
Taking the limit of $z_{f} \rightarrow 0$, we obtain
\bqa && \frac{1}{4} \frac{1}{L} \sum_{k} \frac{1}{\sqrt{- 2 t \cos k + m^{2} - i \varphi}} \coth \Bigg( \frac{\beta}{2} \sqrt{- 2 t \cos k + m^{2} - i \varphi} \Bigg) = \Bigg( 1 - \frac{t^{4}}{2 [m^{2}]^{2} \Big(2 [m^{2}]^{2} + u\Big)} \Bigg) \frac{i \varphi}{2 u} , \eqa
which is nothing but an RG-improved mean-field equation, where the self-interaction constant is renormalized in the one-loop level. This demonstration shows that the present dual holographic effective field theory generalizes the standard mean-field theory in a non-perturbative way, where renormalization effects of both all the coupling constants and dual order parameter fields are taken into account. More precisely, Eq. (\ref{IR_Boundary_Condition_Nonperturbativeness}) takes into account quantum corrections of the all-loop order, where not only interaction renormalizations described by their RG flows but also self-energy corrections up to the $1/N^{f}$ order are self-consistently resumed through the technique of recursive RG transformations. We recall that $f$ is the iteration number of recursive RG transformations. In particular, the self-consistent intertwined renormalization structure between all the coupling functions and the dual order-parameter field is responsible for the description of non-perturbative physics \cite{Holographic_Description_RG_GR}. Of course, this statement does not mean that this theoretical framework is exact. Instead, it is given by the large $N$ limit that allows us to neglect quantum fluctuations of order parameter fields. In this respect we claim that our dual holographic effective field theory serves as a novel mean-field theory beyond the conventional mean-field theoretical framework.

\section{Emergent geometry in recursive renormalization group transformations} \label{Equivalence_Lattice_Continuum}

\subsection{Equivalence between two effective partition functions}

In this section we show the equivalence between
\bqa && Z = \int D \Phi(\mathbf{x}) D \varphi(z) D m^{2}(z) D t(z) D u(z) \delta \Big(t(0) - t_{0}\Big) \delta \Big(m^{2}(0) - m_{0}^{2} + 2 t_{0}\Big) \delta \Big(u(0) - u_{0}\Big) \nn && \delta \Bigg( \partial_{z} m^{2}(z) + \frac{2 m^{2}(z) [t(z)]^{2}}{a \Big(2 [m^{2}(z)]^{2} + u(z)\Big) } \Bigg) \delta \Bigg( \partial_{z} t(z) + \frac{1}{a} t(z) - \frac{m^{2}(z) [t(z)]^{2}}{a\Big(2 [m^{2}(z)]^{2} + u(z)\Big)} - \frac{i}{2} \partial_{z} \varphi(z) \Bigg) \nn && \delta \Bigg( \partial_{z} u(z) + \frac{1}{a} u(z) - \frac{u(z) [t(z)]^{4}}{2 a [m^{2}(z)]^{2} \Big(2 [m^{2}(z)]^{2} + u(z)\Big)} \Bigg) \nn && \exp\Big[ - \int d^{2} \mathbf{x} \Big\{ \Big(\partial_{\tau} \Phi(\mathbf{x})\Big)^{2} + t(z_{f}) \Big( \partial_{x} \Phi(\mathbf{x}) \Big)^{2} + \Big( m^{2}(z_{f}) - i \varphi(z_{f}) - 2 t(z_{f}) \Big) [\Phi(\mathbf{x})]^{2} \Big\} \nn && - \int d^{2} \mathbf{x} \frac{1}{2u(0)} [\varphi(0)]^{2} \nn && - \int_{0}^{z_{f}} d z \int d^{2} \mathbf{x} \Big\{ \frac{a}{2 u(z)} \Big( \partial_{z} \varphi(z)\Big)^{2} + \frac{u(z)}{2 a \Big(2 [m^{2}(z)]^{2} + u(z)\Big)} + \frac{m(z)}{4 a} \Big\} \Big] \label{Equivalence_Lattice_Continuum_ELFT} \eqa
and
\bqa && Z = \int D \phi(\mathbf{x}) D \tilde{\varphi}(z) D g_{\tau\tau}(z) D g_{xx}(z) \delta\Big(g^{\tau\tau}(0) - 1\Big) \delta\Big(g^{xx}(0) - v_{\phi}^{2} \Big) \nn && \delta\Big\{\partial_{z} g^{\tau\tau}(z) - g^{\tau\tau}(z) \Big(\partial_{\tau} \partial_{\tau'} G_{\mathbf{x}\mathbf{x}'}[g_{\mu\nu}(z),\tilde{\varphi}(z)]\Big)_{\mathbf{x}' \rightarrow \mathbf{x}} g^{\tau'\tau}(z) \Big\} \nn && \delta\Big\{\partial_{z} g^{xx}(z) - g^{xx}(z) \Big(\partial_{x} \partial_{x'} G_{\mathbf{x}\mathbf{x}'}[g_{\mu\nu}(z),\tilde{\varphi}(z)]\Big)_{\mathbf{x}' \rightarrow \mathbf{x}} g^{x'x}(z) \Big\} \nn && \exp\Big[ - \int d^{2} \mathbf{x} \sqrt{g(z_{f})} \Big\{ g^{\tau\tau}(z_{f}) [\partial_{\tau} \phi(\mathbf{x})]^{2} + g^{xx}(z_{f}) [\partial_{x} \phi(\mathbf{x})]^{2} + [m^{2} - i \tilde{\varphi}(z_{f})] [\phi(\mathbf{x})]^{2} \Big\} \nn && - \int d^{2} \mathbf{x} \sqrt{g(0)} \frac{1}{2u} [\tilde{\varphi}(0)]^{2} \nn && - \int_{0}^{z_{f}} d z \int d^{2} \mathbf{x} \sqrt{g(z)} \Big\{ \frac{1}{2u} [\partial_{z} \tilde{\varphi}(z)]^{2} + \frac{1}{2 \kappa} \Big( R(z) - 2 \Lambda \Big) \Big\} \Big] \label{Holography_Geometry} \eqa
up to a normalization constant. Here, the first partition function is given by the continuum approximation of the lattice field theory Eq. (\ref{Partition_Function_Coupling_RG_Flow}). In the second partition function we neglect the spacetime dependence of both the metric tensor and the collective dual field variable, compared with the first partition function. To avoid confusion due to the use of the same letter, we introduce the tilde symbol to the collective dual field variable in the second partition function. The second partition function has to be supported by the Green's function
\bqa && \Big\{- g^{\tau\tau}(z) \partial_{\tau}^{2} - g^{xx}(z) \partial_{x}^{2} + \frac{1}{\epsilon} [m^{2} - i \tilde{\varphi}(z)] \Big\} G_{\mathbf{x}\mathbf{x}'}[g_{\mu\nu}(z),\tilde{\varphi}(z)] = \frac{1}{\sqrt{g(z)}} \delta^{(2)}(\mathbf{x}-\mathbf{x}') . \label{Green_Ft_Geometry}  \eqa
In these two partition functions $\mathbf{x} = (\tau, x)$ represents a two-dimensional spacetime coordinate.

We show term-by-term correspondences between Eqs. (\ref{Equivalence_Lattice_Continuum_ELFT}) and (\ref{Holography_Geometry}) in subsection IV.B. More precisely, we suggest correspondence equations for not only UV and IR boundary conditions but also the bulk effective actions. Equivalence of the UV boundary conditions guarantees that both quantum field theories are same in the beginning. Correspondence of the bulk effective actions gives rise to an equation to relate these two dual scalar order-parameter fields with each other. Equivalence between the IR boundary conditions results in equations between the coupling functions of hopping integral, renormalized mass, and effective interaction in Eq. (\ref{Equivalence_Lattice_Continuum_ELFT}) and the metric tensor fields in Eq. (\ref{Holography_Geometry}). Finally, we enforce equivalence equations between the RG flows of the coupling functions in Eq. (\ref{Equivalence_Lattice_Continuum_ELFT}) and those of the metric tensor fields in Eq. (\ref{Holography_Geometry}). As a result, we find complete correspondence equations between all microscopic parameters of both effective actions. Simplifying the RG flow equations for the metric tensor fields in subsection IV.C, where translational symmetry is preserved for the ground-state manifold, we represent the RG flow equation of the metric tensor in terms of the running mass parameter in subsection IV.D, which completes proof for the equivalence between two emergent holographic partition functions of Eqs. (\ref{Equivalence_Lattice_Continuum_ELFT}) and (\ref{Holography_Geometry}).

\subsection{Correspondences}

%
%

%
%

It is straightforward to make correspondences between all the terms in these two partition functions. First, we have the correspondence between two UV effective actions
\bqa && \int d^{2} \mathbf{x} \frac{1}{2u(0)} [\varphi(0)]^{2} \Longleftrightarrow \int d^{2} \mathbf{x} \sqrt{g(0)} \frac{1}{2u} [\tilde{\varphi}(0)]^{2} . \eqa
As a result, we obtain two equations of
\bqa && \frac{1}{u(0)} = \frac{\sqrt{g(0)}}{u}
%
%
\eqa
and
\bqa && \varphi(0) = \tilde{\varphi}(0) . \eqa
The first equation gives
\bqa && u_{0} = v_{\phi} u \eqa
between two bare interaction parameters.
It is also trivial to see the correspondence between the hopping integral and the initial velocity as follows
\bqa && t(0) = g^{xx}(0) , \eqa
which results in
\bqa && t_{0} = v_{\phi}^{2} . \eqa
We have one more relation between the bare mass parameters, given by
\bqa && m^{2}(0) - 2 t(0) = m_{0}^{2} - 2 t_{0} = m^{2} . \eqa

%
%

Second, we have the correspondence between two bulk effective actions in the following way
\bqa && \int_{0}^{z_{f}} d z \int d^{2} \mathbf{x} \Big\{ \frac{a}{2 u(z)} \Big( \partial_{z} \varphi(z)\Big)^{2} + \frac{u(z)}{2 a \Big(2 [m^{2}(z)]^{2} + u(z)\Big)} + \frac{m(z)}{4 a} \Big\} \nn && \Longleftrightarrow \int_{0}^{z_{f}} d z \int d^{2} \mathbf{x}  \sqrt{g(z)} \Big\{ \frac{1}{2u} [\partial_{z} \tilde{\varphi}(z)]^{2} + \frac{1}{2 \kappa} \Big( R(z) - 2 \Lambda \Big) \Big\} . \eqa
This correspondence gives rise to two equations
\bqa && \frac{a}{2 u(z)} \Big( \partial_{z} \varphi(z)\Big)^{2} = \sqrt{g(z)} \frac{1}{2u} [\partial_{z} \tilde{\varphi}(z)]^{2} \eqa
for the dynamics of the bulk dual field variable and
\bqa && \frac{u(z)}{2 a \Big(2 [m^{2}(z)]^{2} + u(z)\Big)} + \frac{m(z)}{4 a} + \mathcal{C}[m(z);z] = \frac{1}{2 \kappa} \sqrt{g(z)} \Big( R(z) - 2 \Lambda \Big) \label{Bulk_Correspondence} \eqa
for the vacuum-fluctuation energy, respectively. In the second equation we introduced a function $\mathcal{C}[m(z);z]$, which depends on the renormalized mass parameter only. Below, we show that this functional can result from the difference between two RG schemes.

%
%

%
%

%
%

Third, we have the correspondence between two IR effective actions as follows
\bqa && \int d^{2} \mathbf{x} \Big\{ \Big(\partial_{\tau} \Phi(\mathbf{x})\Big)^{2} + t(z_{f}) \Big( \partial_{x} \Phi(\mathbf{x}) \Big)^{2} + \Big( m^{2}(z_{f}) - i \varphi(z_{f}) - 2 t(z_{f}) \Big) [\Phi(\mathbf{x})]^{2} \Big\} \nn && \Longleftrightarrow \int d^{2} \mathbf{x} \sqrt{g(z_{f})} \Big\{ g^{\tau\tau}(z_{f}) [\partial_{\tau} \phi(\mathbf{x})]^{2} + g^{xx}(z_{f}) [\partial_{x} \phi(\mathbf{x})]^{2} + [m^{2} - i \tilde{\varphi}(z_{f})] [\phi(\mathbf{x})]^{2} \Big\} . \eqa
We point out that the RG scheme with the metric tensor preserves the relativistic invariance with the Euclidean signature in two dimensions while the Lorentz invariance is explicitly broken in the Kadanoff block-spin transformation. To make a correspondence between these two IR effective actions, it is necessary to consider
\bqa && \Phi(\mathbf{x}) = \Big( \sqrt{g(z_{f})} g^{\tau\tau}(z_{f}) \Big)^{\frac{1}{2}} \phi(\mathbf{x}) . \eqa
As a result, we obtain the following correspondence
\bqa && \int d^{2} \mathbf{x} \sqrt{g(z_{f})} \Big\{ g^{\tau\tau}(z_{f}) [\partial_{\tau} \phi(\mathbf{x})]^{2} + g^{\tau\tau}(z_{f})  t(z_{f}) [\partial_{x} \phi(\mathbf{x})]^{2} + g^{\tau\tau}(z_{f}) \Big( m^{2}(z_{f}) - i \varphi(z_{f}) - 2 t(z_{f}) \Big) [\phi(\mathbf{x})]^{2} \Big\} \nn && \Longleftrightarrow \int d^{2} \mathbf{x} \sqrt{g(z_{f})} \Big\{ g^{\tau\tau}(z_{f}) [\partial_{\tau} \phi(\mathbf{x})]^{2} + g^{xx}(z_{f}) [\partial_{x} \phi(\mathbf{x})]^{2} + [m^{2} - i \tilde{\varphi}(z_{f})] [\phi(\mathbf{x})]^{2} \Big\} . \eqa
This relation results in two equations
\bqa && g^{\tau\tau}(z_{f}) t(z_{f}) = g^{xx}(z_{f}) \eqa
for the kinetic energy and
\bqa && g^{\tau\tau}(z_{f}) \Big(m^{2}(z_{f}) - i \varphi(z_{f}) - 2 t(z_{f})\Big) = m^{2} - i \tilde{\varphi}(z_{f}) \label{IR_Boundary_Matching} \eqa
for the mass term.

%
%

Finally, we have the correspondence between the RG $\beta-$functions of the coupling constants and the RG-flow equations of the metric tensor
\bqa && \partial_{z} m^{2}(z) = - \frac{2 m^{2}(z) [t(z)]^{2}}{a \Big(2 [m^{2}(z)]^{2} + u(z)\Big) } , ~~~~~ \partial_{z} t(z) = - \frac{1}{a} t(z) + \frac{m^{2}(z) [t(z)]^{2}}{a\Big(2 [m^{2}(z)]^{2} + u(z)\Big)} + \frac{i}{2} \partial_{z} \varphi(z) , \nn && \partial_{z} u(z) = - \frac{1}{a} u(z) + \frac{u(z) [t(z)]^{4}}{2 a [m^{2}(z)]^{2} \Big(2 [m^{2}(z)]^{2} + u(z)\Big)} \nn && \Longleftrightarrow \partial_{z} g^{\tau\tau}(z) = g^{\tau\tau}(z) \Big(\partial_{\tau} \partial_{\tau'} G_{\mathbf{x}\mathbf{x}'}[g_{\mu\nu}(z),\tilde{\varphi}(z)]\Big)_{\mathbf{x}' \rightarrow \mathbf{x}} g^{\tau'\tau}(z) , \nn && \partial_{z} g^{xx}(z) = g^{xx}(z) \Big(\partial_{x} \partial_{x'} G_{\mathbf{x}\mathbf{x}'}[g_{\mu\nu}(z),\tilde{\varphi}(z)]\Big)_{\mathbf{x}' \rightarrow \mathbf{x}} g^{x'x}(z) . \eqa

%
%

%
%

In summary, we must have correspondence equations between
\bqa && t(z) , ~~~ m^{2}(z) , ~~~ u(z) , ~~~ \varphi(z) , ~~~ a \eqa
and
\bqa && g^{\tau\tau}(z) , ~~~ g^{xx}(z) , ~~~ \tilde{\varphi}(z) , ~~~ \epsilon . \eqa
We recall that $a$ ($\epsilon$) is an energy scale of the recursive Kadanoff block-spin transformations (the recursive RG transformations with the metric tensor). It seems that the number of the corresponding variables does not match between these correspondences. However, we point out that $t(z)$ is related with $\varphi(z)$ in the RG $\beta-$function. Indeed, we have the following relation
\bqa && \frac{a}{2 u(z)} \Big( \partial_{z} \varphi(z)\Big)^{2} = - \frac{2 a}{u(z)} \Bigg( \partial_{z} t(z) + \frac{1}{a} t(z) - \frac{m^{2}(z) [t(z)]^{2}}{a\Big(2 [m^{2}(z)]^{2} + u(z)\Big)} \Bigg)^{2} , \eqa
performing the path integral with respect to $\varphi(z)$ in Eq. (\ref{Equivalence_Lattice_Continuum_ELFT}), where the $\delta-$function constraint gives rise to the replacement between $\partial_{z} \varphi(z)$ and $\partial_{z} t(z)$. As a result, we have four independent parameters, where their correspondences are determined by four equations: two RG-flow equations
\bqa && \partial_{z} m^{2}(z) = - \frac{2 m^{2}(z) [t(z)]^{2}}{a \Big(2 [m^{2}(z)]^{2} + u(z)\Big) } , ~~~~~ \partial_{z} u(z) = - \frac{1}{a} u(z) + \frac{u(z) [t(z)]^{4}}{2 a [m^{2}(z)]^{2} \Big(2 [m^{2}(z)]^{2} + u(z)\Big)} \nn && \Longleftrightarrow \partial_{z} g^{\tau\tau}(z) = g^{\tau\tau}(z) \Big(\partial_{\tau} \partial_{\tau'} G_{\mathbf{x}\mathbf{x}'}[g_{\mu\nu}(z),\tilde{\varphi}(z)]\Big)_{\mathbf{x}' \rightarrow \mathbf{x}} g^{\tau'\tau}(z) , \nn && \partial_{z} g^{xx}(z) = g^{xx}(z) \Big(\partial_{x} \partial_{x'} G_{\mathbf{x}\mathbf{x}'}[g_{\mu\nu}(z),\tilde{\varphi}(z)]\Big)_{\mathbf{x}' \rightarrow \mathbf{x}} g^{x'x}(z) , \eqa
one kinetic-energy equation
\bqa && g^{\tau\tau}(z) t(z) = g^{xx}(z) , \eqa
and the bulk-field correspondence equation
\bqa && \frac{a}{2 u(z)} \Big( \partial_{z} \varphi(z)\Big)^{2} = - \frac{2 a}{u(z)} \Bigg( \partial_{z} t(z) + \frac{1}{a} t(z) - \frac{m^{2}(z) [t(z)]^{2}}{a\Big(2 [m^{2}(z)]^{2} + u(z)\Big)} \Bigg)^{2} = \sqrt{g(z)} \frac{1}{2u} [\partial_{z} \tilde{\varphi}(z)]^{2} . \eqa
The correspondence between both UV and IR boundary conditions have been presented above. In particular, the matching condition Eq. (\ref{IR_Boundary_Matching}) for the IR boundary guarantees the equivalence for the solution of the order parameter field, given by the second-order differential equation.

%
%

%
%
%

\subsection{RG flow equations for the metric tensor}

To represent the metric tensor in terms of running mass and interaction parameters, we first simplify the RG-flow equations of the metric tensor. We point out that the Green's function is
\bqa && G(q,i\Omega) = \frac{1}{\sqrt{g(z)}} \frac{1}{g^{\tau\tau}(z) \Omega^{2} + g^{xx}(z) q^{2} + \frac{1}{\epsilon} [m^{2} - i \tilde{\varphi}(z)]} \label{Green_Ft_Geometry_Momentum} \eqa
in the frequency-momentum space. Then, the RG-flow equation is given by
\bqa && \partial_{z} \ln g^{\tau\tau}(z) = \frac{1}{\sqrt{g(z)}} \int \frac{d \Omega}{2\pi} \int \frac{d q}{2\pi} \frac{g^{\tau\tau}(z) \Omega^{2}}{g^{\tau\tau}(z) \Omega^{2} + g^{xx}(z) q^{2} + \frac{1}{\epsilon} [m^{2} - i \tilde{\varphi}(z)]} \eqa
for the time component and
\bqa && \partial_{z} \ln g^{xx}(z) = \frac{1}{\sqrt{g(z)}} \int \frac{d \Omega}{2\pi} \int \frac{d q}{2\pi} \frac{g^{xx}(z) q^{2}}{g^{\tau\tau}(z) \Omega^{2} + g^{xx}(z) q^{2} + \frac{1}{\epsilon} [m^{2} - i \tilde{\varphi}(z)]} \eqa
for the space component, respectively.

One can simplify these expressions further as follows
\bqa && \partial_{z} \ln g^{\tau\tau}(z) = \int \frac{d y}{2\pi} \int \frac{d x}{2\pi} \frac{y^{2}}{y^{2} + x^{2} + \frac{1}{\epsilon} [m^{2} - i \tilde{\varphi}(z)]} , \nn && \partial_{z} \ln g^{xx}(z) = \int \frac{d x}{2\pi} \int \frac{d y}{2\pi} \frac{y^{2}}{x^{2} + y^{2} + \frac{1}{\epsilon} [m^{2} - i \tilde{\varphi}(z)]} . \eqa
As a result, we obtain \cite{Integral}
\bqa && \partial_{z} \ln g^{\tau\tau}(z) = \frac{4}{\epsilon} \Big( \int_{0}^{\Lambda} \frac{d w}{2\pi} \frac{w^{2}}{\sqrt{w^{2} + 1}} \Big) \Big( \int_{0}^{\infty} \frac{d z}{2\pi} \frac{1}{z^{2} + 1} \Big) [m^{2} - i \tilde{\varphi}(z)] \label{Solution_Metric_Time_Component} \eqa
for the time component of the metric tensor, where the space component is given by
\bqa && \frac{g^{\tau\tau}(z)}{g^{\tau\tau}(0)} = \frac{g^{xx}(z)}{g^{xx}(0)} . \label{Solution_Metric_Space_Component} \eqa
Here, the upper cutoff $\Lambda$ is proportional to $z_{f}$.

%
%

\subsection{Consistency between RG $\beta-$functions and metric evolution equations}

%
%
%
%

Resorting to $g^{\tau\tau}(z) t(z) = g^{xx}(z)$ and $\frac{g^{\tau\tau}(z)}{g^{\tau\tau}(0)} = \frac{g^{xx}(z)}{g^{xx}(0)}$, we find that the hopping integral does not run, given by
\bqa && t(z) = \frac{g^{xx}(0)}{g^{\tau\tau}(0)} . \eqa
This coincides with the initial condition for the hopping integral, given by $t_{0} = v_{\phi}^{2}$.

%
%

Next, we reformulate the RG $\beta-$functions of the mass and interaction parameters. We introduce an effective coupling constant as
\bqa && \mathcal{U}(z) \equiv \frac{u(z)}{ 2 [m^{2}(z)]^{2} } . \eqa
%
%
%
%
%
%
%
%
%
%
%
%
Then, one can reformulate the RG-flow equation of the self-interaction parameter in terms of this effective coupling constant. Considering the constant hopping integral, we find that the effective interaction coefficient is given by the mass parameter as follows
\bqa && \mathcal{U}(z) = \mathcal{U}(0) \frac{[m^{2}(0)]^{2}}{[m^{2}(z)]^{2}} \exp\Big\{ - \frac{1}{a} z + \frac{t_{0}^{2}}{8} \Big( \frac{1}{[m^{2}(z)]^{2}} - \frac{1}{[m^{2}(0)]^{2}} \Big) \Big\} . \label{Effective_Interaction_RG} \eqa
Inserting this expression into the RG $\beta-$function of the mass parameter, we obtain
\bqa && a \partial_{z} [m^{2}(z)]^{2} = - \frac{2 t_{0}^{2} [m^{2}(z)]^{2}}{[m^{2}(z)]^{2} + \mathcal{U}(0) [m^{2}(0)]^{2} \exp\Big\{ - \frac{1}{a} z + \frac{t_{0}^{2}}{8} \Big( \frac{1}{[m^{2}(z)]^{2}} - \frac{1}{[m^{2}(0)]^{2}} \Big) \Big\}} , \eqa
which implies non-perturbative nature of this RG framework.

%
%
%

%
%

The final task is to represent the RG-flow equation of the metric tensor in terms of the running coupling constants. Taking one more derivative with respect to $z$ for the evolution equation of the metric tensor, we obtain
\bqa && \partial_{z}^{2} \ln g^{\tau\tau}(z) = \frac{4}{\epsilon} \Big( \int_{0}^{\Lambda} \frac{d w}{2\pi} \frac{w^{2}}{\sqrt{w^{2} + 1}} \Big) \Big( \int_{0}^{\infty} \frac{d z}{2\pi} \frac{1}{z^{2} + 1} \Big) [ - i \partial_{z} \tilde{\varphi}(z)] . \eqa
The bulk-field correspondence equation is rewritten as
\bqa && i \partial_{z} \tilde{\varphi}(z) = \Big( \frac{a \sqrt{t_{0}} u g^{\tau\tau}(z)}{2 [m^{2}(z)]^{2} \mathcal{U}(z)}\Big)^{1/2} i \partial_{z} \varphi(z) . \eqa
%
%
%
Inserting this expression into the above, we obtain
\bqa && [g^{\tau\tau}(z)]^{-1/2} \partial_{z}^{2} \ln g^{\tau\tau}(z) = \frac{4}{\epsilon} \Big( \int_{0}^{\Lambda} \frac{d w}{2\pi} \frac{w^{2}}{\sqrt{w^{2} + 1}} \Big) \Big( \int_{0}^{\infty} \frac{d z}{2\pi} \frac{1}{z^{2} + 1} \Big) \Big( \frac{a \sqrt{t_{0}} u}{2 [m^{2}(z)]^{2} \mathcal{U}(z)}\Big)^{1/2} [ - i \partial_{z} \varphi(z)] . \label{Metric_RG_Flow_Correspondence} \eqa

We recall that the RG $\beta-$function of the hopping integral is
\bqa && i \partial_{z} \varphi(z) = \frac{2}{a} t_{0} - \frac{t_{0}^{2}}{a m^{2}(z)\Big(1 + \mathcal{U}(z) \Big)} . \eqa
Using Eq. (\ref{Effective_Interaction_RG}), we rewrite this $\beta-$function as
\bqa && i a \partial_{z} \varphi(z) = 2 t_{0} - \frac{t_{0}^{2} m^{2}(z)}{[m^{2}(z)]^{2} + \mathcal{U}(0) [m^{2}(0)]^{2} \exp\Big\{ - \frac{1}{a} z + \frac{t_{0}^{2}}{8} \Big( \frac{1}{[m^{2}(z)]^{2}} - \frac{1}{[m^{2}(0)]^{2}} \Big) \Big\}} . \eqa
Inserting this expression into Eq. (\ref{Metric_RG_Flow_Correspondence}), we represent the RG flow of the metric tensor in terms of the running mass parameter in the following way
\bqa && [g^{\tau\tau}(z)]^{-1/2} \partial_{z}^{2} \ln g^{\tau\tau}(z) \nn && = - \frac{4}{a \epsilon} \Big( \int_{0}^{\Lambda} \frac{d w}{2\pi} \frac{w^{2}}{\sqrt{w^{2} + 1}} \Big) \Big( \int_{0}^{\infty} \frac{d z}{2\pi} \frac{1}{z^{2} + 1} \Big) \Bigg( \frac{a \sqrt{t_{0}} u}{2 \mathcal{U}(0) [m^{2}(0)]^{2} \exp\Big\{ - \frac{1}{a} z + \frac{t_{0}^{2}}{8} \Big( \frac{1}{[m^{2}(z)]^{2}} - \frac{1}{[m^{2}(0)]^{2}} \Big) \Big\}} \Bigg)^{1/2} \nn && \Bigg( 2 t_{0} - \frac{t_{0}^{2} m^{2}(z)}{[m^{2}(z)]^{2} + \mathcal{U}(0) [m^{2}(0)]^{2} \exp\Big\{ - \frac{1}{a} z + \frac{t_{0}^{2}}{8} \Big( \frac{1}{[m^{2}(z)]^{2}} - \frac{1}{[m^{2}(0)]^{2}} \Big) \Big\}} \Bigg) . \label{Metric_Mass_Relation} \eqa
This expression completes our proof.

Finally, we reconsider Eq. (\ref{Bulk_Correspondence}). Introducing $\mathcal{U}(z) \equiv \frac{u(z)}{ 2 [m^{2}(z)]^{2} }$ into Eq. (\ref{Bulk_Correspondence}), we obtain
\bqa && \frac{\mathcal{U}(z)}{1 + \mathcal{U}(z)} + \frac{m(z)}{2} + \mathcal{C}[m(z);z] = \frac{a}{\kappa} \sqrt{g(z)} \Big( R(z) - 2 \Lambda \Big) . \nonumber \eqa
Considering that $\mathcal{U}(z)$ is a functional of the renormalized mass parameter $m(z)$, given by Eq. (\ref{Effective_Interaction_RG}), we find that the left hand side can be represented by purely a running mass parameter with the extra-dimensional space coordinate $z$. Recalling that the metric tensor is given by the running mass parameter only, given by Eq. (\ref{Metric_Mass_Relation}), we realize that the right hand side can be also expressed in terms of only the running mass parameter. As a result, the difference between the vacuum energy, $\mathcal{C}[m(z);z]$ depends on $m(z)$ only with $z$ explicitly at most.

\section{Emergence of AdS$_{3}$ geometry at the quantum critical point} \label{QCP_AdS3}

\subsection{Dual holographic effective field theory in the metric formulation}

We recall the effective partition function Eq. (\ref{Holography_Geometry})
\bqa && Z = \int D \phi(\mathbf{x}) D \tilde{\varphi}(z) D g_{\tau\tau}(z) D g_{xx}(z) \delta\Big(g^{\tau\tau}(0) - 1\Big) \delta\Big(g^{xx}(0) - v_{\phi}^{2} \Big) \nn && \delta\Big\{\partial_{z} g^{\tau\tau}(z) - g^{\tau\tau}(z) \Big(\partial_{\tau} \partial_{\tau'} G_{\mathbf{x}\mathbf{x}'}[g_{\mu\nu}(z),\tilde{\varphi}(z)]\Big)_{\mathbf{x}' \rightarrow \mathbf{x}} g^{\tau'\tau}(z) \Big\} \nn && \delta\Big\{\partial_{z} g^{xx}(z) - g^{xx}(z) \Big(\partial_{x} \partial_{x'} G_{\mathbf{x}\mathbf{x}'}[g_{\mu\nu}(z),\tilde{\varphi}(z)]\Big)_{\mathbf{x}' \rightarrow \mathbf{x}} g^{x'x}(z) \Big\} \nn && \exp\Big[ - \int d^{2} \mathbf{x} \sqrt{g(z_{f})} \Big\{ g^{\tau\tau}(z_{f}) [\partial_{\tau} \phi(\mathbf{x})]^{2} + g^{xx}(z_{f}) [\partial_{x} \phi(\mathbf{x})]^{2} + [m^{2} - i \tilde{\varphi}(z_{f})] [\phi(\mathbf{x})]^{2} \Big\} \nn && - \int d^{2} \mathbf{x} \sqrt{g(0)} \frac{1}{2u} [\tilde{\varphi}(0)]^{2} \nn && - \int_{0}^{z_{f}} d z \int d^{2} \mathbf{x} \sqrt{g(z)} \Big\{ \frac{1}{2u} [\partial_{z} \tilde{\varphi}(z)]^{2} + \frac{1}{2 \kappa} \Big( R(z) - 2 \Lambda \Big) \Big\} \Big] . \nonumber \eqa
Here, the Green's function for the RG flow of the metric tensor in the $\delta-$function constraint is given by Eq. (\ref{Green_Ft_Geometry})
\bqa && \Big\{- g^{\tau\tau}(z) \partial_{\tau}^{2} - g^{xx}(z) \partial_{x}^{2} + \frac{1}{\epsilon} [m^{2} - i \tilde{\varphi}(z)] \Big\} G_{\mathbf{x}\mathbf{x}'}[g_{\mu\nu}(z),\tilde{\varphi}(z)] = \frac{1}{\sqrt{g(z)}} \delta^{(2)}(\mathbf{x}-\mathbf{x}') . \nonumber \eqa
%
%
Inserting the Fourier transformation of the Green's function Eq. (\ref{Green_Ft_Geometry_Momentum}) into the RG flow of the metric tensor, we obtain Eq. (\ref{Solution_Metric_Time_Component})
%
%
%
%
%
\bqa && \partial_{z} \ln g^{\tau\tau}(z) = \frac{4}{\epsilon} \Big( \int_{0}^{\Lambda} \frac{d w}{2\pi} \frac{w^{2}}{\sqrt{w^{2} + 1}} \Big) \Big( \int_{0}^{\infty} \frac{d z}{2\pi} \frac{1}{z^{2} + 1} \Big) [m^{2} - i \tilde{\varphi}(z)] \nonumber \eqa
for the time component, where the space component is given by Eq. (\ref{Solution_Metric_Space_Component})
\bqa && g^{xx}(z) = \frac{g^{xx}(0)}{g^{\tau\tau}(0)} g^{\tau\tau}(z) . \nonumber \eqa

%
%

%
%

The ``dynamics" of the static dual order parameter field is given by the RG flow in the above emergent curved spacetime background. Performing the variation of the effective action with respect to $\tilde{\varphi}(z)$ and $\partial_{z} \tilde{\varphi}(z)$, we obtain the Lagrange equation of motion in the extra-dimensional curved spacetime
\bqa && \frac{\partial_{z} \sqrt{g(z)}}{u} \partial_{z} \tilde{\varphi}(z) + \frac{\sqrt{g(z)}}{u} \partial_{z}^{2} \tilde{\varphi}(z) = 0 . \eqa
%
%
%
%
%
It is straightforward to solve this second-order differential equation, the solution of which is given by
\bqa && \tilde{\varphi}(z) = \tilde{\varphi}(0) + [\partial_{z} \tilde{\varphi}(z)]_{z = 0} \int_{0}^{z} d z' \sqrt{\frac{g(0)}{g(z')}} . \eqa
%
%
There are two unknown coefficients, which have to be determined by UV and IR boundary conditions. These boundary conditions should be also derived from the effective field theory itself, as discussed before.

The bulk effective action gives rise to both UV and IR boundary terms as follows
\bqa && \int_{0}^{z_{f}} d z \int d^{2} \mathbf{x} \sqrt{g(z)} \frac{1}{2u} [\partial_{z} \tilde{\varphi}(z)]^{2} \nn && = \int_{0}^{z_{f}} d z \int d^{2} \mathbf{x} \Big\{ - \frac{1}{4u} \Big( \sqrt{\frac{g_{xx}(z)}{g_{\tau\tau}(z)}} \partial_{z} g_{\tau\tau}(z) + \sqrt{\frac{g_{\tau\tau}(z)}{g_{xx}(z)}} \partial_{z} g_{xx}(z) \Big) \tilde{\varphi}(z) [\partial_{z} \tilde{\varphi}(z)] + \frac{\sqrt{g(z)}}{2u} \tilde{\varphi}(z) [- \partial_{z}^{2} \tilde{\varphi}(z)] \Big\} \nn && + \int d^{2} \mathbf{x} \Big\{ \frac{\sqrt{g(z_{f})}}{2u} \tilde{\varphi}(z_{f}) [\partial_{z} \tilde{\varphi}(z)]_{z = z_{f}} - \frac{\sqrt{g(0)}}{2u} \tilde{\varphi}(0) [\partial_{z} \tilde{\varphi}(z)]_{z = 0} \Big\} , \eqa
where the integration-by-parts has been utilized. Then, we obtain the IR effective action
\bqa && \mathcal{S}_{IR}[\tilde{\varphi}(z_{f})] = \int d^{2} \mathbf{x} \sqrt{g(z_{f})} \Big\{ g^{\tau\tau}(z_{f}) [\partial_{\tau} \phi(x)]^{2} + g^{xx}(z_{f}) [\partial_{x} \phi(x)]^{2} + [m^{2} - i \tilde{\varphi}(z_{f})] \phi^{2}(x) + \frac{1}{2u} \tilde{\varphi}(z_{f}) [\partial_{z} \tilde{\varphi}(z)]_{z = z_{f}} \Big\} \nn \eqa
and the UV boundary action
\bqa && \mathcal{S}_{UV}[\tilde{\varphi}(0)] = \int d^{2} \mathbf{x} \sqrt{g(0)} \Big\{ \frac{1}{2u} [\tilde{\varphi}(0)]^{2} - \frac{1}{2u} \tilde{\varphi}(0) [\partial_{z} \tilde{\varphi}(z)]_{z = 0} \Big\} , \eqa
respectively.

%
%

Performing the variation of the UV boundary action with respect to $\tilde{\varphi}(0)$ and $[\partial_{z} \tilde{\varphi}(z)]_{z = 0}$, we obtain the UV boundary condition
\bqa && \frac{1}{4u} \Big( \sqrt{\frac{g_{xx}(0)}{g_{\tau\tau}(0)}} [\partial_{z} g_{\tau\tau}(z)]_{z = 0} + \sqrt{\frac{g_{\tau\tau}(0)}{g_{xx}(0)}} [\partial_{z} g_{xx}(z)]_{z = 0} \Big) \tilde{\varphi}(0) + \frac{\sqrt{g(0)}}{u} \tilde{\varphi}(0) = 0 . \eqa
As a result, we find
\bqa && \tilde{\varphi}(0) = 0 , \eqa
consistent with the case of the Kadanoff block-spin transformation in section \ref{Holography_Lattice}. The solution is
\bqa && \tilde{\varphi}(z) = [\partial_{z} \tilde{\varphi}(z)]_{z = 0} \int_{0}^{z} d z' \sqrt{\frac{g(0)}{g(z')}} , \eqa
where $[\partial_{z} \tilde{\varphi}(z)]_{z = 0}$ is determined by the IR boundary condition.

%
%
%

Performing the variation of the IR boundary effective action with respect to $\tilde{\varphi}(z_{f})$ and $[\partial_{z} \tilde{\varphi}(z)]_{z = z_{f}}$, we obtain the IR boundary condition
\bqa && \frac{1}{4u} \Big( \sqrt{\frac{g_{xx}(z_{f})}{g_{\tau\tau}(z_{f})}} [\partial_{z} g_{\tau\tau}(z)]_{z = z_{f}} + \sqrt{\frac{g_{\tau\tau}(z_{f})}{g_{xx}(z_{f})}} [\partial_{z} g_{xx}(z)]_{z = z_{f}} \Big) \tilde{\varphi}(z_{f}) + i \sqrt{g(z_{f})} \Big\langle \phi^{2}(x) \Big\rangle = 0 . \eqa
Resorting to the Green's function Eq. (\ref{Green_Ft_Geometry_Momentum}), we rewrite this expression as
\bqa && \int \frac{d \Omega}{2 \pi} \int \frac{d q}{2 \pi} \frac{1}{g^{\tau\tau}(z_{f}) \Omega^{2} + g^{xx}(z_{f}) q^{2} + [m^{2} - i \tilde{\varphi}(z_{f})]} \nn && = \frac{1}{2u} \Big( \sqrt{\frac{g_{xx}(z_{f})}{g_{\tau\tau}(z_{f})}} [\partial_{z} g_{\tau\tau}(z)]_{z = z_{f}} + \sqrt{\frac{g_{\tau\tau}(z_{f})}{g_{xx}(z_{f})}} [\partial_{z} g_{xx}(z)]_{z = z_{f}} \Big) [i \tilde{\varphi}(z_{f})] , \eqa
where $\tilde{\varphi}(z_{f})$ is related with  $[\partial_{z} \tilde{\varphi}(z)]_{z = 0}$ in the following way $\tilde{\varphi}(z_{f}) = [\partial_{z} \tilde{\varphi}(z)]_{z = 0} \int_{0}^{z_{f}} d z' \sqrt{\frac{g(0)}{g(z')}}$. Here, $g(z') = g_{\tau\tau}(z') g_{xx}(z')$ is the determinant of the metric tensor.
%
%
%
%

It is straightforward to solve this equation, the solution of which is given by
\bqa && i \tilde{\varphi}(z_{f}) = 4 \Big( \int_{0}^{\Lambda} \frac{d w}{2 \pi} \frac{1}{\sqrt{w^{2} + 1}} \Big) \Big( \int_{0}^{\infty} \frac{d z}{2 \pi} \frac{1}{z^{2} + 1} \Big) \frac{2u}{[\partial_{z} \ln g(z)]_{z = z_{f}}} , \eqa
where $\Lambda$ is the same cutoff introduced before.
%
%
%
As a result, we find the solution of the collective dual field variable
\bqa && i \tilde{\varphi}(z) = 4 \Big( \int_{0}^{\Lambda} \frac{d w}{2 \pi} \frac{1}{\sqrt{w^{2} + 1}} \Big) \Big( \int_{0}^{\infty} \frac{d z}{2 \pi} \frac{1}{z^{2} + 1} \Big) \frac{2u}{[\partial_{z} \ln g(z)]_{z = z_{f}} \int_{0}^{z_{f}} d z' \sqrt{\frac{g(0)}{g(z')}}} \int_{0}^{z} d z' \sqrt{\frac{g(0)}{g(z')}} , \label{Solution_Order_Parameter_Geometry} \eqa
which describes how the mean-field value of the order parameter field evolves as a function of an energy scale.

\subsection{Emergent geometry}

%
%

Finally, we are ready to deduce the emergent geometry in this dual holographic effective field theory. Inserting the solution of the order parameter field Eq. (\ref{Solution_Order_Parameter_Geometry}) into the RG flow of the metric tensor Eq. (\ref{Solution_Metric_Time_Component}), we obtain
\bqa && \partial_{z} \ln g^{\tau\tau}(z) = \frac{4}{\epsilon} \Big( \int_{0}^{\Lambda} \frac{d w}{2\pi} \frac{w^{2}}{\sqrt{w^{2} + 1}} \Big) \Big( \int_{0}^{\infty} \frac{d z}{2\pi} \frac{1}{z^{2} + 1} \Big) \Bigg\{ m^{2} \nn && - 4 \Big( \int_{0}^{\Lambda} \frac{d w}{2 \pi} \frac{1}{\sqrt{w^{2} + 1}} \Big) \Big( \int_{0}^{\infty} \frac{d z}{2 \pi} \frac{1}{z^{2} + 1} \Big) \frac{2u}{[\partial_{z} \ln g(z)]_{z = z_{f}} \int_{0}^{z_{f}} d z' \sqrt{\frac{g(0)}{g(z')}}} \int_{0}^{z} d z' \sqrt{\frac{g(0)}{g(z')}} \Bigg\} . \eqa
%
%
The RG flow in the one-loop level is given by
\bqa && [\partial_{z} \ln g^{\tau\tau}(z)]_{z = 0} = \frac{4}{\epsilon} \Big( \int_{0}^{\Lambda} \frac{d w}{2\pi} \frac{w^{2}}{\sqrt{w^{2} + 1}} \Big) \Big( \int_{0}^{\infty} \frac{d z}{2\pi} \frac{1}{z^{2} + 1} \Big) m^{2} , \eqa
where the limit of $z \rightarrow 0$ has been taken into account. On the other hand, the full RG flow is given by
\bqa && [\partial_{z} \ln g^{\tau\tau}(z)]_{z = z_{f}} = \frac{4}{\epsilon} \Big( \int_{0}^{\Lambda} \frac{d w}{2\pi} \frac{w^{2}}{\sqrt{w^{2} + 1}} \Big) \Big( \int_{0}^{\infty} \frac{d z}{2\pi} \frac{1}{z^{2} + 1} \Big) \Bigg\{ m^{2} \nn && - 4 \Big( \int_{0}^{\Lambda} \frac{d w}{2 \pi} \frac{1}{\sqrt{w^{2} + 1}} \Big) \Big( \int_{0}^{\infty} \frac{d z}{2 \pi} \frac{1}{z^{2} + 1} \Big) \frac{2u}{[\partial_{z} \ln g(z)]_{z = z_{f}} } \Bigg\} , \eqa
where $z = z_{f}$ has been considered.

Inserting $g(z) = g_{\tau\tau}(z) g_{xx}(z)$ into the above expression and using $g^{xx}(z) = \frac{g^{xx}(0)}{g^{\tau\tau}(0)} g^{\tau\tau}(z)$, we find the solution of this RG flow at $z = z_{f}$
%
%
%
%
\bqa && [\partial_{z} \ln g^{\tau\tau}(z)]_{z = z_{f}} = 2 \Bigg[ \frac{1}{\epsilon} \Big( \int_{0}^{\Lambda} \frac{d w}{2\pi} \frac{w^{2}}{\sqrt{w^{2} + 1}} \Big) \Big( \int_{0}^{\infty} \frac{d z}{2\pi} \frac{1}{z^{2} + 1} \Big) m^{2} \nn && \pm \sqrt{\Big\{ \frac{1}{\epsilon} \Big( \int_{0}^{\Lambda} \frac{d w}{2\pi} \frac{w^{2}}{\sqrt{w^{2} + 1}} \Big) \Big( \int_{0}^{\infty} \frac{d z}{2\pi} \frac{1}{z^{2} + 1} \Big) m^{2} \Big\}^{2} + \frac{4}{\epsilon} \Big( \int_{0}^{\Lambda} \frac{d w}{2\pi} \frac{w^{2}}{\sqrt{w^{2} + 1}} \Big) \Big( \int_{0}^{\Lambda} \frac{d w}{2 \pi} \frac{1}{\sqrt{w^{2} + 1}} \Big) \Big( \int_{0}^{\infty} \frac{d z}{2 \pi} \frac{1}{z^{2} + 1} \Big)^{2} u} \Bigg] . \nn \label{Metric_Solution_IR} \eqa
Accordingly, the renormalized value of the order parameter field is
\bqa && i \tilde{\varphi}(z_{f}) = - 4 \Big( \int_{0}^{\Lambda} \frac{d w}{2 \pi} \frac{1}{\sqrt{w^{2} + 1}} \Big) \Big( \int_{0}^{\infty} \frac{d z}{2 \pi} \frac{1}{z^{2} + 1} \Big) \frac{u}{[\partial_{z} \ln g^{\tau\tau}(z)]_{z = z_{f}}} . \eqa
The resulting effective mass parameter is given by
\bqa && M^{2}(z_{f}) \equiv m^{2} - i \tilde{\varphi}(z_{f}) = m^{2} + 4 \Big( \int_{0}^{\Lambda} \frac{d w}{2 \pi} \frac{1}{\sqrt{w^{2} + 1}} \Big) \Big( \int_{0}^{\infty} \frac{d z}{2 \pi} \frac{1}{z^{2} + 1} \Big) \frac{u}{[\partial_{z} \ln g^{\tau\tau}(z)]_{z = z_{f}}} . \eqa

%
%

First, we pick up the minus sign in Eq. (\ref{Metric_Solution_IR}), given by
\bqa && [\partial_{z} \ln g^{\tau\tau}(z)]_{z = z_{f}} = 2 \Bigg[ \frac{1}{\epsilon} \Big( \int_{0}^{\Lambda} \frac{d w}{2\pi} \frac{w^{2}}{\sqrt{w^{2} + 1}} \Big) \Big( \int_{0}^{\infty} \frac{d z}{2\pi} \frac{1}{z^{2} + 1} \Big) m^{2} \nn && - \sqrt{\Big\{ \frac{1}{\epsilon} \Big( \int_{0}^{\Lambda} \frac{d w}{2\pi} \frac{w^{2}}{\sqrt{w^{2} + 1}} \Big) \Big( \int_{0}^{\infty} \frac{d z}{2\pi} \frac{1}{z^{2} + 1} \Big) m^{2} \Big\}^{2} + \frac{4}{\epsilon} \Big( \int_{0}^{\Lambda} \frac{d w}{2\pi} \frac{w^{2}}{\sqrt{w^{2} + 1}} \Big) \Big( \int_{0}^{\Lambda} \frac{d w}{2 \pi} \frac{1}{\sqrt{w^{2} + 1}} \Big) \Big( \int_{0}^{\infty} \frac{d z}{2 \pi} \frac{1}{z^{2} + 1} \Big)^{2} u} \Bigg] . \nn \eqa
%
%
Taking the limit of $\epsilon \rightarrow 0$, we obtain
\bqa && [\partial_{z} \ln g^{\tau\tau}(z)]_{z = z_{f}} \approx - 4 \Big( \int_{0}^{\infty} \frac{d z}{2\pi} \frac{1}{z^{2} + 1} \Big) \Big( \int_{0}^{\Lambda} \frac{d w}{2 \pi} \frac{1}{\sqrt{w^{2} + 1}} \Big) \frac{u}{m^{2}} . \label{RG_Flow_Metric_IR_QCP} \eqa
Then, the order parameter field takes
\bqa && i \tilde{\varphi}(z_{f}) \approx m^{2} . \eqa
As a result, we are at the quantum critical point
\bqa && M^{2}(z_{f}) \equiv m^{2} - i \tilde{\varphi}(z_{f}) \approx 0 . \eqa

It is straightforward to solve Eq. (\ref{RG_Flow_Metric_IR_QCP}). The time component of the metric tensor is give by
\bqa && g^{\tau\tau}(z_{f}) \approx g^{\tau\tau}(z) \exp\Big\{ - 4 \Big( \int_{0}^{\infty} \frac{d z}{2\pi} \frac{1}{z^{2} + 1} \Big) \Big( \int_{0}^{\Lambda} \frac{d w}{2 \pi} \frac{1}{\sqrt{w^{2} + 1}} \Big) \frac{u}{m^{2}} (z_{f} - z) \Big\} . \eqa
Accordingly, the space component is
\bqa && g^{xx}(z_{f}) \approx v_{\phi}^{2} g^{\tau\tau}(z) \exp\Big\{ - 4 \Big( \int_{0}^{\infty} \frac{d z}{2\pi} \frac{1}{z^{2} + 1} \Big) \Big( \int_{0}^{\Lambda} \frac{d w}{2 \pi} \frac{1}{\sqrt{w^{2} + 1}} \Big) \frac{u}{m^{2}} (z_{f} - z) \Big\} . \eqa
This metric tensor is nothing but the AdS$_{3}$ geometry if the coordinate of the extra-dimensional space is appropriately scaled. As a result, we confirm the emergence of the AdS$_{3}$ metric at the quantum critical point of the transverse-field Ising model.

%
%

Next, we consider the positive sign in Eq. (\ref{Metric_Solution_IR}), given by
\bqa && [\partial_{z} \ln g^{\tau\tau}(z)]_{z = z_{f}} = 2 \Bigg[ \frac{1}{\epsilon} \Big( \int_{0}^{\Lambda} \frac{d w}{2\pi} \frac{w^{2}}{\sqrt{w^{2} + 1}} \Big) \Big( \int_{0}^{\infty} \frac{d z}{2\pi} \frac{1}{z^{2} + 1} \Big) m^{2} \nn && + \sqrt{\Big\{ \frac{1}{\epsilon} \Big( \int_{0}^{\Lambda} \frac{d w}{2\pi} \frac{w^{2}}{\sqrt{w^{2} + 1}} \Big) \Big( \int_{0}^{\infty} \frac{d z}{2\pi} \frac{1}{z^{2} + 1} \Big) m^{2} \Big\}^{2} + \frac{4}{\epsilon} \Big( \int_{0}^{\Lambda} \frac{d w}{2\pi} \frac{w^{2}}{\sqrt{w^{2} + 1}} \Big) \Big( \int_{0}^{\Lambda} \frac{d w}{2 \pi} \frac{1}{\sqrt{w^{2} + 1}} \Big) \Big( \int_{0}^{\infty} \frac{d z}{2 \pi} \frac{1}{z^{2} + 1} \Big)^{2} u} \Bigg] . \nn \eqa
Taking the limit of $\epsilon \rightarrow 0$, we obtain
\bqa && [\partial_{z} \ln g^{\tau\tau}(z)]_{z = z_{f}} \approx \frac{4}{\epsilon} \Big( \int_{0}^{\Lambda} \frac{d w}{2\pi} \frac{w^{2}}{\sqrt{w^{2} + 1}} \Big) \Big( \int_{0}^{\infty} \frac{d z}{2\pi} \frac{1}{z^{2} + 1} \Big) m^{2} , \label{RG_Flow_Metric_IR_Gapped} \eqa
which results in
\bqa && i \tilde{\varphi}(z_{f}) = - \epsilon \Big( \int_{0}^{\Lambda} \frac{d w}{2 \pi} \frac{1}{\sqrt{w^{2} + 1}} \Big) \Big( \int_{0}^{\Lambda} \frac{d w}{2\pi} \frac{w^{2}}{\sqrt{w^{2} + 1}} \Big)^{-1} \frac{u}{m^{2}} . \eqa
As a result, we are in a gapped phase, given by the effective mass parameter
\bqa && M^{2}(z_{f}) \equiv m^{2} - i \tilde{\varphi}(z_{f}) = m^{2} + \epsilon \Big( \int_{0}^{\Lambda} \frac{d w}{2 \pi} \frac{1}{\sqrt{w^{2} + 1}} \Big) \Big( \int_{0}^{\Lambda} \frac{d w}{2\pi} \frac{w^{2}}{\sqrt{w^{2} + 1}} \Big)^{-1} \frac{u}{m^{2}} . \eqa

Solving Eq. (\ref{RG_Flow_Metric_IR_Gapped}), we obtain
\bqa && g^{\tau\tau}(z_{f}) \approx g^{\tau\tau}(z) \exp\Big\{ \frac{1}{\epsilon} \Big( \int \frac{d w}{2\pi} \frac{w^{2}}{\sqrt{w^{2} + 1}} \Big) \Big( \int \frac{d z}{2\pi} \frac{1}{z^{2} + 1} \Big) m^{2} (z_{f} - z) \Big\} , \eqa
which shows divergence of the metric tensor in the $z_{f} \rightarrow \infty$ limit. This RG flow indicates a run-away behavior in the gapped phase.

\section{Discussion: Fermion-boson duality in geometry} \label{Discussion}

Recently, we revisited the quantum phase transition in the transverse-field Ising model and translated it in terms of emergent geometry \cite{Holographic_Description_Entanglement_Entropy}. There, we started from the Majorana-fermion representation, where the transverse-field Ising model turns into a p-wave superconductor model of spinless fermions, given by \cite{Kitaev_Model}
\bqa && Z = \int D \psi_{i}(\tau) \exp\Big[ - \int_{0}^{\beta} d \tau \sum_{i = 1}^{L} \Big\{ \psi_{i}^{\dagger}(\tau) \partial_{\tau} \psi_{i}(\tau) - J \lambda \psi_{i}^{\dagger}(\tau) \tau_{3} \psi_{i}(\tau) - J \psi_{i}^{\dagger}(\tau) (\tau_{3} - i \tau_{2}) \psi_{i-1}(\tau) \Big\} \Big] . \label{Majorana_TFI} \eqa
Here, $\psi_{i}(\tau)$ is the Nambu spinor for superconductivity and $\tau_{l}$ with $l = 1, 2, 3$ is the Pauli matrix acting on the Nambu-spinor basis. $J$ is the ferromagnetic exchange interaction between Ising spins and $J \lambda$ is the transverse magnetic field. It is well known that this effective lattice model shows a quantum phase transition from a p-wave topologically nontrivial BCS superconducting phase in $\lambda < \lambda_{c}$ to a p-wave topologically trivial BEC superconducting state in $\lambda > \lambda_{c}$, where $\lambda_{c}$ is the quantum critical point, described by two copies of Majorana fermions with $c = 1/2$ \cite{Kitaev_Model}. Here, $c$ is the central charge.

Implementing the Kadanoff block-spin transformation in a recursive way, we obtained an effective field theory \cite{Holographic_Description_Entanglement_Entropy}
\bqa && Z = \int D \psi_{i}(\tau) \delta\Big(J(0) - J\Big) \delta\Big(\Phi(0) - J \lambda\Big) \delta \Big(a \partial_{z} J(z) + J(z) - \frac{2 [J(z)]^{2}}{\Phi(z)}\Big) \delta\Big(a \partial_{z} \Phi(z) \Big) \nn && \exp\Big[ - \int_{0}^{\beta} d \tau \sum_{i = 1}^{L} \Big\{ \psi_{i}^{\dagger}(\tau) \partial_{\tau} \psi_{i}(\tau) - \Phi(z_{f}) \psi_{i}^{\dagger}(\tau) \tau_{3} \psi_{i}(\tau) - J(z_{f}) \psi_{i}^{\dagger}(\tau) (\tau_{3} - i \tau_{2}) \psi_{i-1}(\tau) \Big\} \nn && + \frac{1}{a} \int_{0}^{z_{f}} d z \int_{0}^{\beta} d \tau \sum_{i = 1}^{L} \frac{1}{\beta} \ln \Big\{ 2 \cosh \Big( \frac{\beta}{2} \Phi(z) \Big) \Big\} \Big] , \label{Holography_Majorana_TFI} \eqa
which manifests the RG flow of the effective action. Here, both the ferromagnetic exchange interaction parameter $J(0) = J$ and the transverse magnetic field $\Phi(0) = J \lambda$ show their RG flows, given by $a \partial_{z} J(z) = - J(z) + \frac{2 [J(z)]^{2}}{\Phi(z)}$ and $a \partial_{z} \Phi(z) = 0$ and shown in the $\delta-$function constraints, respectively. Their RG flows terminate at $z = z_{f}$, and they appear in the IR effective action $\mathcal{S}_{IR} = \int_{0}^{\beta} d \tau \sum_{i = 1}^{L} \Big\{ \psi_{i}^{\dagger}(\tau) \partial_{\tau} \psi_{i}(\tau) - \Phi(z_{f}) \psi_{i}^{\dagger}(\tau) \tau_{3} \psi_{i}(\tau) - J(z_{f}) \psi_{i}^{\dagger}(\tau) (\tau_{3} - i \tau_{2}) \psi_{i-1}(\tau) \Big\}$. The bulk action $\mathcal{S}_{Bulk} = \frac{1}{a} \int_{0}^{z_{f}} d z \int_{0}^{\beta} d \tau \sum_{i = 1}^{L} \frac{1}{\beta} \ln \Big\{ 2 \cosh \Big( \frac{\beta}{2} \Phi(z) \Big) \Big\}$ describes vacuum fluctuations of even-site fermions.

Based on this effective field theory, we translated the quantum phase transition of Eq. (\ref{Majorana_TFI}) in terms of emergent geometry at zero temperature \cite{Holographic_Description_Entanglement_Entropy}. Taking the continuum approximation for Eq. (\ref{Holography_Majorana_TFI}), we could extract out the metric tensor to describe each superconducting phase. As expected, the quantum critical point is given by the AdS$_{3}$ metric. On the other hand, the topologically trivial superconducting phase is described by AdS$_{2}$ $\times$ $R$. An interesting point is that the topologically nontrivial superconducting state is characterized by the appearance of a horizon at some length along the extra-dimensional space, where the Ricci curvature diverges and the emergent spacetime terminates. To confirm the validity of this horizon geometry, we calculated the holographic entanglement entropy referred to as Ryu-Takayanagi formula \cite{Entanglement_Entropy_Ryu_Takayanagi_I,Entanglement_Entropy_Ryu_Takayanagi_II}. It turns out that this holographic entanglement entropy coincides with that of the corresponding field theory \cite{Entanglement_Entropy_Calabrese_Cardy_I,Entanglement_Entropy_Calabrese_Cardy_II} not only at the quantum critical point but also in the topologically nontrivial superconducting phase. Unfortunately, the holographic entanglement entropy in the topologically trivial superconducting state did not match that of the field theory, which implies that the AdS$_{2}$ $\times$ $R$ geometry may be valid only in the IR limit.

In this study we also found the AdS$_{3}$ geometry at the quantum critical point of the bosonic representation. On the other hand, we got a divergent geometry of the $z_{f} \rightarrow \infty$ limit in the paramagnetic phase corresponding to the topologically trivial superconducting state in the Majorana-fermion representation. At present, it is not clear how the fermion-boson duality is realized in terms of geometry.

\section{Conclusion}

In conclusion, we proposed a prescription for a dual holographic description, based on recursive Wilsonian renormalization group (RG) transformations. First, we introduced an effective dual holographic action, where the RG flow of the metric tensor manifests in the IR effective action through the emergent extra-dimensional space. The bulk effective action gives rise to the RG flow of the dual order-parameter field in the large $N$ limit, identified with the Callan-Symanzik equation of a one-particle Green's function and supported by UV and IR boundary conditions. It is the intertwined renormalization structure between the metric tensor and the dual order-parameter field that allows a non-perturbative description. In particular, we explicitly demonstrated that the IR boundary condition generalizes the conventional mean-field theory, taking into account renormalization effects through the RG flow of the metric tensor in a self-consistent way. Second, we presented a version of condensed matter physics for a dual holographic description. Based on recursive Kadanoff block-spin transformations in two spacetime dimensions, we obtained an effective dual holographic theory, which manifests the RG flow of the effective action through the emergent extra-dimensional space. The RG flow of all coupling constants appears in the IR effective action, the RG flow of which is realized in the extra-dimensional space. The bulk effective action describes the RG flow of the order-parameter field in the presence of the RG flow of the interaction vertices, supported by both UV and IR boundary conditions. In particular, we showed that the IR boundary condition extends the conventional mean-field theory, which replaces the bare interaction parameters with their renormalized coupling constants in a self-consistent way. Finally, we showed the equivalence between the first and the second dual holographic prescriptions. Resorting to term-by-term matching between these two dual holographic effective actions, we reformulated the RG flow of the metric tensor in terms of the running coupling constants of the microscopic lattice model. Based on this equivalence, we verified that the AdS$_{3}$ geometry emerges at the quantum critical point of the transverse-field Ising model.

\begin{acknowledgments}
K.-S. Kim was supported by the Ministry of Education, Science, and Technology (No. 2011-0030046) of the National Research Foundation of Korea (NRF) and by TJ Park Science Fellowship of the POSCO TJ Park Foundation. K.-S. Kim appreciates fruitful discussions with Shinsei Ryu and his hospitality during the sabbatical leave.
\end{acknowledgments}


\begin{thebibliography}{9}
\bibitem{Integral} Changing the integral variable as $x = \sqrt{y^{2} + \frac{m^{2} - i \tilde{\varphi}(z)}{\epsilon}} z$ in the first equation, we obtain \bqa && \partial_{z} \ln g^{\tau\tau}(z) = 2 \int_{-\infty}^{\infty} \frac{d y}{2\pi} \frac{y^{2}}{\sqrt{y^{2} + \frac{m^{2} - i \tilde{\varphi}(z)}{\epsilon}}} \int_{0}^{\infty} \frac{d z}{2\pi} \frac{1}{z^{2} + 1} . \nonumber \eqa Considering the change of the integral variable as $y = \sqrt{\frac{m^{2} - i \tilde{\varphi}(z)}{\epsilon}} w$, we obtain \bqa && \partial_{z} \ln g^{\tau\tau}(z) = \frac{4}{\epsilon} \Big( \int_{0}^{\infty} \frac{d w}{2\pi} \frac{w^{2}}{\sqrt{w^{2} + 1}} \Big) \Big( \int_{0}^{\infty} \frac{d z}{2\pi} \frac{1}{z^{2} + 1} \Big) [m^{2} - i \tilde{\varphi}(z)] . \nonumber \eqa Since the $w$ integral is divergent, we introduce a UV cut-off scale $\Lambda$, resulting in Eq. (\ref{Solution_Metric_Time_Component}). This cut-off scale defines our UV quantum field theory.
\end{thebibliography}
\end{document}